\documentclass[usenatbib]{mnras}

\usepackage{newtxtext,newtxmath}

\usepackage[T1]{fontenc}
\usepackage{ae,aecompl}

\usepackage{graphicx}	
\usepackage{amsmath}	
\usepackage{dblfloatfix}
\usepackage{siunitx}
\usepackage[dvipsnames]{xcolor}

\usepackage{caption}
\usepackage{subcaption}

\usepackage{cleveref}
\crefname{section}{§}{§§}

\title[Rotating Viscous Dissipation]{Viscous Dissipation and Dynamics in Simulations of Rotating, Stratified Plane-layer Convection}

\author[Lance et al.]{
Simon R. W. Lance$^{1}$ \thanks{E-mail: s.lance@exeter.ac.uk}, Laura K. Currie$^{2} \thanks{E-mail: laura.currie@durham.ac.uk}$ and Matthew K. Browning$^{1}$ \thanks{E-mail: m.k.m.browning@exeter.ac.uk} 
\\
$^{1}$Physics and Astronomy Department, University of Exeter, Stocker Rd, Exeter, EX4 4QL, UK\\
$^{2}$Department of Mathematical Sciences, Durham University, Durham, DH1 3LE, UK\\
}

\pubyear{2024}
\date{Accepted 2024 January 16. Received 2023 December 22; in original form 2023 July 4}

\begin{document}
\label{firstpage}
\pagerange{\pageref{firstpage}--\pageref{lastpage}}
\maketitle

\defcitealias{Hewitt75}{HMW75}
\defcitealias{CurrieBrowning17}{CB17}

\begin{abstract}
Convection in stars and planets must be
maintained against viscous and Ohmic dissipation. Here, we present the first systematic investigation of viscous dissipation in simulations of rotating, density-stratified plane layers of convection.  Our simulations consider an anelastic ideal gas, and employ the
open-source code Dedalus.  We
demonstrate that when the convection is sufficiently vigorous, the integrated dissipative
heating tends towards a value that is independent of viscosity or thermal diffusivity, but depends on the imposed luminosity and the stratification.  We show that knowledge of the dissipation provides a bound on the magnitude of the kinetic energy flux in the convection zone. In our non-rotating cases with simple flow fields, much of the dissipation occurs near the highest possible temperatures, and the kinetic energy flux approaches this bound.  In the rotating cases, although the total integrated dissipation is
similar, it is much more uniformly distributed (and locally balanced by work  against the stratification), with a consequently smaller
kinetic energy flux. The heat transport in our rotating simulations is in good agreement with results previously obtained for 3D Boussinesq convection, and approaches the predictions of diffusion-free theory. 

\end{abstract}




\section{Introduction}

Convection occurs in the interior of every main-sequence star and in many 
planets, and must be maintained against a finite amount of viscous and
Ohmic dissipation.  In a steady state, the dynamics and the dissipation are
therefore linked; constraints on one yield constraints on the other.

Many authors have explored this link.  For example, the widely-employed
theory of Rayleigh-B\'enard convection developed by \citet{Grossmann_Lohse_2000}
relies on the exact relationship between viscous dissipation and heat
transport \citep{Shraiman_Siggia_2000}; in their model, the heat transport
depends crucially on whether the viscous and thermal dissipation occur
primarily in the bulk of the convective domain or in the boundary layers
that form at its top and base.  \citet{Jones_et_al_2022} have recently explored
an extension of this theory to the density-stratified case, with the spatial
distribution of the dissipation again playing a vital role.  In the stellar
context, \citet{Anders_et_al_2022} have shown that the magnitude of the
dissipation within a convection zone strongly influences the amount of
convective overshooting into adjacent stable layers. The form
and magnitude of the dissipation is likewise crucial in a variety of
efforts to go ``beyond mixing length theory'' (MLT) (e.g., \citealt {Kupka_et_al_2022}, \citealt{Canuto_1997}, \citealt{Viallet13}, \citealt {Meakin_Arnett_2010},
\citealt{Arnett_et_al_2015}). In the Sun, where the form and magnitude of
convective flows in the deep convection zone are currently the subject of
much debate \citep[e.g.,][]{Vasil_et_al_2021}, the total dissipation may provide
important constraints on the flows \citep{Ginet_1994}. Ohmic
dissipation in particular is thought to limit the depth of zonal winds in
Jupiter (\citealt{Liu_et_al_2008}, \citealt{Kaspi_et_al_2018}, \citealt{Kaspi_et_al_2020}), may constrain
magnetism in the interiors of low-mass stars \citep{Browning_et_al_2016}, and
could influence the radii of hot Jupiters \citep{Batygin_Stevenson_2010}.

The purpose of this paper is to provide new constraints on the magnitude
and spatial distribution of the viscous dissipation that may be occurring in
stellar convection zones.  Few prior works have systematically
investigated this in the astrophysically-relevant case of a gas with
density and temperature stratification, and none have done so when rotation
is also present.  Here we study this issue within one of the simplest
possible systems that captures convection, rotation, and stratification, by
conducting a series of hydrodynamic simulations of stratified (anelastic)
convection in a rotating Cartesian domain, situated at a fixed latitude.
The vast majority of the simulations presented here are 2D, although we compare some of these
results to a very small number of 3D simulations.  Many elements that are
important in real stars --- including, crucially, magnetic fields --- are
thus absent here.  However, this setup has the great advantage that it
allows us to sample parameter regimes that would be difficult or impossible
to probe in equivalent detail in a full 3D spherical geometry.

In particular, we are able to assess how the dissipation scales with
luminosity, rotation rate, and stratification in the
limit where the diffusivities are small
(i.e., when the convective supercriticality is high). In what follows,
we argue that in this regime the dissipation rate (integrated over the
convection zone) depends only on the luminosity and the stratification, and
is (at fixed supercriticality) independent of rotation.  
However, the spatial distribution of this dissipation --- and with it, many
other aspects of the dynamics --- does depend on rotation, as detailed below. 

In the remainder of this
introduction, we summarise prior bounds on the viscous and Ohmic
dissipation, and describe how our work extends these.  In Section 2 we
detail our simulation setup. In Section 3 we provide a brief, qualitative overview of the dynamics in our simulations.  In Section 4 we examine the magnitude and spatial
distribution of the dissipation in these simulations, and  how these scale with the convective driving, the stratification,
and the rotational influence.  In Section 5 we explore the links between the
dissipation, dynamics, and heat transport. We show there that knowledge of the dissipation provides novel constraints
on the kinetic energy flux. We
close in Section 6 with a summary of our results and their
possible astrophysical implications.  

\subsection{Overview of prior work: bounds and constraints on dissipative heating}

In the interior of a star, the microphysical diffusion of momentum, heat,
or magnetic fields is typically very small compared to other physical processes, so that the relevant non-dimensional numbers (e.g., the Reynolds, Rayleigh, and magnetic
Reynolds numbers) are usually very large (e.g., \citealt{Kulsrud_2005}, \citealt{Brun_Browning_2017}, \citealt{Jermyn_et_al_2022}).
This need not imply, however, that viscous and Ohmic dissipation are
negligible.  

To place our discussion on a firmer footing, and to highlight some of the
aims of our work, we briefly describe the thermodynamic constraints on the
dissipation here.  More complete discussions can be found in \citet{Hewitt75} (hereafter HMW75), in \cite{Backus_1975}, \cite{Alboussiere_Ricard_2013, Alboussiere_Ricard_2014},
and \cite{Alboussiere_et_al_2022}.

Consider a volume $V$ of convecting fluid with an associated magnetic field
$\boldsymbol{B}$, enclosed by some surface $S$.  Assume this surface is
impenetrable and either stress-free or no-slip, so that the normal
component of the fluid velocity $\boldsymbol{u}$, and either all components
of $\boldsymbol{u}$ or the tangential stress vanish on $S$. The local rate
of change of total energy can be expressed by 
\begin{align}\label{consofE}\frac{\partial}{\partial{t}}\left(\rho{e}+\frac{1}{2}\rho{u}^2\right.&\left.+\frac{B^2}{2\mu_0}-\rho\Psi\right)=-\nabla\cdot\left(\rho\left(e+\frac{1}{2}u^2-\Psi\right)\mathbf{u}\right.\nonumber\\&\left.+\frac{(\mathbf{E}\times\mathbf{B})}{\mu_0}+P\mathbf{u}-\mathbf{\tau}\cdot\mathbf{u}-k\nabla{T}\right)+H\end{align}
with $e$ the fluid's internal energy, $\rho$ its density, $\Psi$
the gravitational potential satisfying $\mathbf{g}=\nabla\Psi$, $P$  the
pressure, $\tau_{ij}$ the contribution to the total stress tensor from
irreversible processes, $k$ the thermal conductivity, $T$ the
temperature, $H$ the rate of internal heat generation (e.g., by nuclear
fusion or radioactive decay) or cooling (e.g., by any processes not included in the conductive term), and $\mu_0$ the
 permeability of free space. We have assumed the MHD approximation holds, so that $\mathbf{E} = -\mathbf{u} \times \mathbf{B} + \eta \nabla \times \mathbf{B}$, where $\eta = 1/(\mu_0 \sigma)$ is the magnetic diffusivity and $\sigma$ the electrical conductivity \citep[e.g.,][]{priest2014_book}.  Physically, the rate of total energy change
 at a point is given by the sum of the net inward flux of energy (the
 divergence terms in eqn. \ref{consofE}) and the rate of internal heat
 generation.

The first global constraint is that total energy is conserved,
but this yields little insight into the magnitude of the
dissipative heating.   Integrating (\ref{consofE}) over $V$ gives
 \begin{equation}\label{Fbal}
\int_Sk\frac{\partial{T}}{\partial{x_i}}\,dS_i+\int_VH\,dV=0,
 \end{equation}
assuming both a steady state and that the electric current, ${\mathbf{j} = (\nabla \times \mathbf{B})/\mu_0}$,
vanishes everywhere outside $V$. Equation (\ref{Fbal}) implies that the net
flux out of $V$ is equal to the total rate of internal heating and cooling. But
dissipative terms do not appear in this equation; viscous and
ohmic heating do not contribute to the overall heat flux.

To constrain the dissipation, we turn
instead to
the internal energy equation, which can be written as:
\begin{equation}
    \rho \left(\frac{\partial e}{\partial t} + (\boldsymbol{u} \cdot \nabla)e  \right) = \nabla \cdot (k\nabla T) - P(\nabla \cdot \boldsymbol{u}) + \tau_{ij}\frac{\partial u_i}{\partial x_j} + \frac{j^2}{\sigma} + H .
    \label{eq:int_E_eq}
\end{equation} Assuming a steady state and integrating over the fluid
volume V it can be shown that
\begin{equation}
    \int_V (\boldsymbol{u} \cdot \nabla)P \mathrm{d}V + \Phi = 0
    \label{eq:Eworkbalance}
\end{equation}
where the total dissipative heating rate, $\Phi$, is defined as
\begin{equation}
    \Phi = \int_V \left(\tau_{ij}\frac{\partial u_i}{\partial x_j} + \frac{j^2}{\sigma} \right)\, \mathrm{d}V.
    \label{eq:Phi}
\end{equation}
The first and second terms inside the integral represent the contributions
due to viscous and Ohmic effects respectively. Equation (\ref{eq:Eworkbalance}) implies that the total
dissipative heating, integrated over the volume, is exactly balanced by the
work done against the background stratification \citep{CurrieBrowning17}.

Equivalently, from the first law of thermodynamics, we have
\begin{equation}
Tds=de-\frac{P}{\rho^2}d\rho
\end{equation}
where $s$ is the specific entropy, implying that
\begin{equation}
    \rho T \left(\frac{\partial s}{\partial t} + (\boldsymbol{u} \cdot
    \nabla)s  \right) = \nabla \cdot (k\nabla T) + \tau_{ij}\frac{\partial u_i}{\partial x_j} + \frac{j^2}{\sigma} + H.
    \label{eq:int_entropy_eq}
\end{equation}
Following  \citetalias{Hewitt75}, we can divide this equation by $T$, and integrate over volume to find
\begin{equation}
  \int_S \frac{k}{T} \nabla T \cdot d\boldsymbol{S} + \int_V k
    \left| \frac{\nabla T}{T} \right|^2 dV +\int_V \frac{1}{T} \left(H + 
    \tau_{ij}\frac{\partial u_i}{\partial x_j} +  \frac{j^2}{\sigma}
    \right) dV = 0.
    \label{eq:entropybound}
\end{equation}

Physically, this equation expresses the fact that there is a flux of
entropy in and out of the domain (first term), and that entropy can be generated in the
bulk by conduction (second term) or by heating within the domain (third
term). If the inward flux of entropy at the bottom is less than the outward flux of entropy out the top ---as occurs if there is a temperature contrast across the domain--- then the difference must be made up by entropy
generation in the convection zone (either by conduction or dissipation).

In \citetalias{Hewitt75} this equation is used to derive an upper limit on the total amount of dissipative heating that can occur in a convective layer of depth $d$. This is given by
\begin{equation}
    E \equiv \frac{\Phi}{L} < \frac{T_0 - T_{\text{top}}}{T_{\text{top}}} ,
    \label{eq:HMWbound}
\end{equation}
where $T_{\text{top}}$ and $T_0$ denote the upper and lower boundary values
of the temperature respectively and $L$ is the luminosity through the layer. This upper limit corresponds to the case in equation
(\ref{eq:entropybound}) where there is negligible entropy generation by
conduction or heating, and where the dissipation occurs at the highest possible
temperature (i.e., at the bottom of the domain). In this case, as discussed in \citetalias{Hewitt75}, the total dissipative heating rate is bounded not by $L$, but by $L d/H_T$, with $H_T$ a suitably-defined temperature scale height.

In general, however, it cannot be assumed that dissipation occurs at the highest possible temperatures. For example, the dissipation could be distributed more uniformly throughout the layer, or it could be concentrated predominantly in boundary layers.  In these situations, $E$ could in principle be much smaller than the upper bound of equation (\ref{eq:HMWbound}).  

Prior simulations have shown that in certain circumstances
convection can approach a version of this bound. \citetalias{Hewitt75} demonstrated that for the specific case of a Boussinesq liquid without
magnetism, the integrated dissipation approached a value of order the bound
at high enough Rayleigh numbers $Ra$ (measuring the ratio of buoyancy
driving to viscous and thermal dissipation).  \cite{JarvisMcKenzie80}
expanded on this by investigating the case of compressible convection in
the infinite Prandtl number $Pr$ (defined as the ratio of viscous to
thermal diffusivities) regime, appropriate for convection
within the Earth's mantle.   \cite{CurrieBrowning17} (hereafter CB17) extended these results to a gas
at finite $Pr$, as appropriate for convection in stellar interiors. In a
series of 2D hydrodynamic simulations without rotation, they found that the total dissipative heating in their calculations obeyed a tighter, but purely empirical bound, specifically (defining
$E\equiv \Phi/L$, with $\Phi$ the total viscous heating and $L$ the luminosity)
\begin{equation}
    E = \frac{d}{\hat{H}_T}
    \label{eq:C&B_lim}
\end{equation}
where 
\begin{equation}
    \hat{H}_T = \frac{H_{T,0}H_{T,{\text{top}}}}{H_{T,z^*}}    
\end{equation}
is a modified thermal scale height involving the scale height at the top
and bottom boundaries ($H_{T,\text{top}}$ and $H_{T,0}$ respectively), and some vertical height $z^*$, defined such that
half of the fluid mass lies above and below $z^*$.  They showed that for
sufficiently high supercriticalities, the value
of $E$ appeared to approach equation (\ref{eq:C&B_lim}) asymptotically. 

Yet not all convective systems actually approach
these upper bounds.  Recently
\citet{Alboussiere_et_al_2022}, studying 2D convection with an unusual equation
of state in which entropy was a function solely of density, found much lower levels of dissipation than suggested by equation (\ref{eq:C&B_lim}) in most cases. They attributed the difference in part to
the different boundary conditions adopted in their work; in particular, they 
showed that for their equation of state, high levels of dissipation
(approaching the bound in eqn. \ref{eq:C&B_lim}) were only realised in
cases with rigid walls \citepalias[as employed in][]{CurrieBrowning17}, and not in those with
periodic boundary conditions. 

Together, these prior results demonstrate that different values of the total dissipative heating are possible in stratified convection.  A central aim of this paper is to provide constraints on how much dissipation actually occurs, for the astrophysically-relevant case of an ideal gas with rotation.

\section{Methodology} 

\subsection{Model Setup}
We model a layer of fluid contained between impermeable, free-slip boundaries at $z = 0$ and $z = d$ and assume that the horizontal boundaries are periodic. Our coordinate system is such that the horizontal coordinates, $x$ and $y$, correspond to longitudinal and latitudinal directions respectively, and the vertical coordinate, $z$, corresponds to the radial axis. In the majority of our simulations, we retain all three components of velocity but assume all variables are independent of $x$, so that those simulations are 2D (however, for a few cases in \S\ref{sec:3dcomparison} we relax this constraint and consider the fully 3D case). Gravity acts in the negative $z$ direction. To drive convection, we impose a flux $F$ at the bottom boundary and fix entropy at the top. Note that for our 2D simulations, $F$ has units of energy per time per length (rather than per area, as in the 3D cases) and $F$ is related to luminosity, $L$, by $F=L/A$ where, in 2D cases, $A$ is the horizontal box length and in 3D cases $A$ is the horizontal cross-sectional area. 

To investigate the effects of rotation on such a convective layer we consider a tilted f-plane, where the rotation vector takes the form $\boldsymbol{\Omega} = (0, \Omega\cos\alpha, \Omega\sin\alpha )$ where $\Omega$ is the rotation rate, and $\alpha$ is the latitude. We conducted cases at $\alpha=90^{\circ}$ and $\alpha=45^{\circ}$; for clarity, in almost all of our discussion below we focus on cases at  $\alpha=90^{\circ}$, which corresponds to a vertical rotation vector, aligned with gravity and representative of polar latitudes on a spherical body. 

To allow for the effects of a density stratification, we use the anelastic equation set under the Lantz-Braginsky-Roberts (LBR) approximation (\citealt{Lantz92}, \citealt{BraginskyRoberst95}). This is expected to be valid when the flows are sufficiently subsonic and the stratification is nearly adiabatic.  We diffuse entropy instead of temperature (see discussions in, e.g., \citealt{Lecoanet14}). We also consider only the hydrodynamical problem, so there is no Lorentz force and all dissipation is viscous.

The governing equations (in dimensionless form) are

\begin{equation}
    \frac{\partial \boldsymbol{u}}{\partial t} + (\boldsymbol{u} \cdot \nabla)\boldsymbol{u} = - \nabla \left(\frac{p}{\bar\rho}\right) + \frac{Ra_F}{Pr} \hat{s} \hat{\boldsymbol{e}}_z -Ta^{\frac{1}{2}}\boldsymbol{\Omega} \times \boldsymbol{u} + \frac{1}{\bar\rho} \frac{\partial}{\partial x_j}\tau_{ij}
    \label{eq:nd_mom_eq}
\end{equation}
\begin{equation}
    \nabla \cdot (\bar{\rho} \boldsymbol{u}) = 0
    \label{eq:nd_continuity_eq}
\end{equation}
\begin{equation}
    Pr\bar{\rho}\bar{T}\left( \frac{\partial s}{\partial t} + \left( \boldsymbol{u} \cdot \nabla \right) s \right) = \nabla \cdot ( \bar{\rho} \bar{T} \nabla s ) + \frac{Pr^2 \theta}{Ra_F}\tau_{ij} \frac{\partial u_i}{\partial x_j}
    \label{eq:nd_entropy_eq}
\end{equation}
where $\boldsymbol{u}=(u,v,w)$ is the fluid velocity, $s$ is the specific entropy, $p$ is the pressure, $\bar \rho$ and $\bar T$ are the reference state density and temperature (defined in (\ref{eq:ref_state}) below) and
\begin{equation}
    \tau_{ij} =  \bar{\rho} \left(\frac{\partial u_i}{\partial x_j} + \frac{\partial u_j}{\partial x_i}  - \frac{2}{3} \delta_{ij} \nabla\cdot\boldsymbol{u} \right).
\end{equation}
These quantities are all dimensionless and were obtained from their dimensional counterparts using $d$ as the characteristic length scale, $d^2/\nu$ as the characteristic time scale (where $\nu$ is the kinematic viscosity), and $\nu/d$ as the characteristic velocity. Specific entropy has characteristic scale $\frac{Fd}{\kappa\rho_0 T_0}$, where $\rho_0$ and $T_0$ are, respectively, the values of the reference state density and temperature at the bottom of the domain and $\kappa$ is the thermal diffusivity. The characteristic scales for $p$, density, and temperature are  $\rho_0\nu^2/d^2$, $\rho_0$, and $T_0$ respectively. Luminosity has scale $FA$, where $A$ has either characteristic scale $d$ in 2D or $d^2$ in 3D.

Equations (\ref{eq:nd_mom_eq}) - (\ref{eq:nd_entropy_eq}) contain several dimensionless parameters defined as follows:
\begin{equation}
    Pr = \frac{\nu}{\kappa}, ~~~~~ Ta = \frac{4 \Omega^2 d^4}{\nu^2}, ~~~~~ Ra_F = \frac{g d^4 F}{\nu \kappa^2 \rho_0 c_{p} T_0}, ~~~~~ \theta = \frac{gd}{c_{p}T_0},
    \label{eq:nd_numbers}
\end{equation}
where $c_p$ is the specific heat capacity at constant pressure and $g$ is the acceleration due to gravity. In this work we take $\nu$, $\kappa$, $c_p$ and $g$ to be constant (i.e., they do not vary with depth). $Pr$ is the Prandtl number and is taken to be unity throughout this study. $Ta$ is the usual Taylor number (quantifying Coriolis forces relative to viscous effects) and $Ra_F$ is a flux-based Rayleigh number. Alongside $Ra_F$ it will also be useful to consider the traditional Rayleigh number defined as $Ra = \frac{gd^3\Delta s}{\kappa \nu}$, where $\Delta s$ is the entropy difference across the layer. $Ta$ and $Ra_F$ will be varied to examine solutions at different rotation rates and at different levels of convective driving. 

The reference state is taken to be a time-independent, hydrostatic, ideal gas given by
\begin{equation}
    \bar{T} = (1 - \theta z), \quad \bar{\rho} = (1 - \theta z)^m,
    \label{eq:ref_state}
\end{equation}
where $m=\frac{3}{2}$ is the polytropic index.

We will refer to the number of density scale heights across our layer, $N_\rho$, to quantify the degree of stratification.  This is defined as
\begin{equation}
    N_{\rho} = \ln \dfrac{\bar\rho_0}{\bar\rho_{\text{top}}} = -m\ln{(1-\theta }).
    \label{eq:Nrho}
\end{equation}

In dimensionless terms the boundary conditions amount to enforcing $\frac{\partial s}{\partial z} = -1$ at $z=0$ and $s=0$ at $z=1$. The impermeable and stress-free boundary conditions become respectively $w=0$ and $\frac{\partial v}{\partial z} = 0 $ on $z=0,1$ (for the 3D simulations in \S \ref{sec:3dcomparison} , we we also have $\frac{\partial u}{\partial z}=0$). 

We solve this system using the pseudo-spectral code Dedalus \citep{Burns_et_al_2020}.
Our simulations at low and moderate $Ra_F$ generally use 256 grid points in the horizontal and 128 in the vertical with an aspect ratio of two (i.e., the box is twice as wide as it is tall); at higher $Ra_F$, higher resolutions were required (with up to 640 grid points in the vertical direction), and in a few of the highest-resolution cases (at $Ta=10^{11}$) we have considered an aspect ratio of 1.075 instead.  The 3D cases in \S \ref{sec:3dcomparison} have an aspect ratio of two. 

The simulations were initialised either by imposing very small entropy perturbations on a motionless base state, or (for some cases at higher $Ra_F$) from an evolved state at lower $Ra_F$.  

It is convenient to specify the buoyancy driving in a convective system by reference to the value of $Ra_F$ at which convection first occurs, the critical Rayleigh number $Ra_c$. To find $Ra_c$, we  constructed an eigenvalue problem (EVP) solver using Dedalus \citep{Burns_et_al_2020}, from which we obtain a grid of growth rates for a given input range of $Ra_F$ and $k_y$ values, where $k_y$ is the horizontal wavenumber. We then used the open-source Eigentools package \citep{oishi_etal2021_eigentools} to find $Ra_c$ (taking into account only those modes that would fit into the finite computational domain). Rotation and stratification both modify the values of $Ra_c$ (see, e.g., \citealt{Chandrasekhar67}, \citealt{Miserski_Tobias_2011}). For the parameters studied here, $Ra_c$ then varies from of order $100$ (for, e.g., cases at $N_{\rho}=4, Ta=0$) to nearly $10^{8}$ (for cases at $Ta=10^{11}$).  

\section{Overview of resulting dynamics}
The convective flows in this system are influenced by rotation, by stratification, and by the level of buoyancy driving.  We have conducted simulations that sample a wide variety of possible states within this multi-dimensional parameter space.  We consider cases ranging from the nearly-Boussinesq limit ($N_{\rho}=0.2$, with a density contrast from top to bottom of only 1.22) up to stronger stratifications with $N_{\rho} = 4$ (density contrast of 55).  The energy passing through the system is quantified by the flux-based Rayleigh number $Ra_F$, as defined above; our simulations sample both laminar flows near convective onset (with $Ra_F$ close to $Ra_c$) and more turbulent states that have $Ra_F \sim 10^6 Ra_c$.  The rotation rate in our simulations is quantified by the Taylor number defined above, which varies between $Ta=10$ and $Ta=10^{11}$. The Ekman number is also commonly used to quantify the influence of rotation relative to viscosity; $Ek=Ta^{-1/2}$, so here varies from $3.16\times10^{-1}$ to $3.16\times10^{-6}$.  We conducted simulations at latitudes of $90^{\circ}$ and $45^{\circ}$, but in almost all the figures below have chosen to focus on cases at $90^{\circ}$ for clarity. (None of the key quantities reported in this paper, or their scalings with $Ra_F$ and $Ta$, appeared to depend significantly on the choice of latitude.)   Table \ref{Table1} lists the input parameters and key derived quantities for a small number of these simulations; the full table is available online.  At each $N_{\rho}$, simulations were performed at a range of logarithmically spaced supercriticalities. For cases performed at a fixed supercriticality (e.g., Figures \ref{fig:E_vs_Ta} and \ref{fig:z_diss_fixed_supercrit_increasing_Ta}), they were instead logarithmically spaced in $Ta$.

\begin{table}
\caption{Input parameters and selected output quantities for example simulations presented in this paper.  Indicated are the number of density scale heights across the layer ($N_{\rho}$), the supercriticality of the simulation $Ra_F/Ra_c$, the critical Rayleigh number $Ra_c$, the Taylor number $Ta$, the convective Rossby number $Ro_c$, the latitude $\alpha$ (for rotating cases only), and the output quantities $E$, and $Nu$. Full machine-readable table available online.   }\label{Table1}
\centering
\begin{tabular}{cccccc|cc}

\hline
$N_{\rho}$ & $Ra_F/Ra_c$ & $Ra_c$ & $Ta$ & $Ro_c$ & $\alpha$ & $E$ & $Nu$  \\
 \hline
\hline 
1.4 & $3.16\times 10^1$ & $6.74\times 10^5$ & $10^8$ & 0.46 & $90^{\circ}$ & 0.82 & 8.61 \\ 
1.4 & $1.78\times 10^2$ & $6.74\times 10^5$ & $10^8$ & 1.09 & $90^{\circ}$ & 0.92 & 15.3\\  
1.4 & $5.62\times 10^2$ & $6.74\times 10^5$ & $10^8$ & 1.95 & $90^{\circ}$ & 0.95 & 19.6 \\ 
1.4 & $10^3$ & $6.74\times 10^5$ & $10^8$ & 2.60 & $90^{\circ}$ & 0.98 & 22.1 \\ 
1.4 & $10^4$ & $6.74\times 10^5$ & $10^8$ & 8.21 & $90^{\circ}$ & 1.05 & 35.0

\end{tabular}
\end{table}

Increasing the rotation rate stabilises the system against convection, increasing the value of $Ra_c$. Thus for simulations at constant $Ra_F$, increasing $Ta$ in isolation would eventually result in a system that no longer convects. In much of our discussions below we therefore choose to compare simulations at varying $Ta$ but constant supercriticality, $Ra_F/Ra_c$. We also quantify rotation using  $\mathrm{Ro}_c = \sqrt{Ra_F/(TaPr)}$ (as in, e.g., \citealt{hindman_etal2020}; see also \citealt{Gilman78}), which assesses the buoyancy driving relative to the Coriolis force (see \citealt{Anders_et_al_2019} for a discussion of how this relates to other measures of rotation).   We sample both rapidly-rotating cases (with some having volume-averaged values of $Ro_c < 1$) and ones in which rotation has little dynamical role ($Ro_c \gg 1$).

\begin{figure}
\centerline{\includegraphics[scale=0.85]{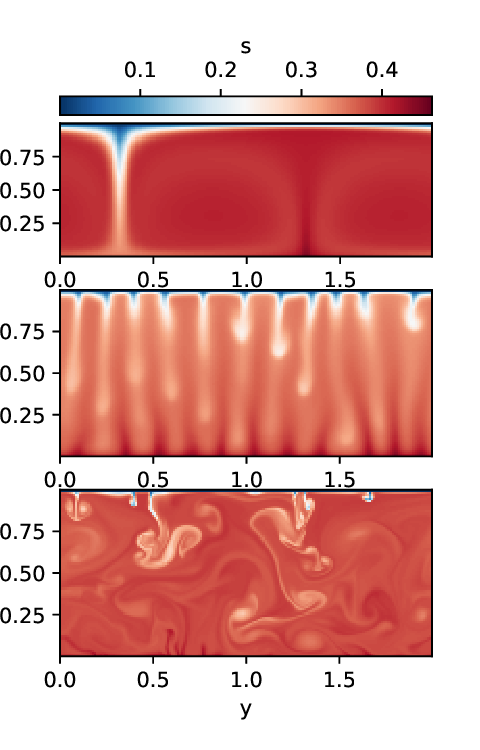}}
    \caption{Specific entropy $s$ in three cases sampling different parameter regimes.  All cases have $N_{\rho}=1.4$; top panel is a non-rotating case ($Ta=0$) with $Ra_F=10^{4} Ra_c$; middle panel has $Ta=10^{8}$ and $Ra_F=56.2 Ra_c$; bottom panel has $Ta=10^{8}$ and $Ra_F = 10^4 Ra_c$. }
    \label{fig:increasing_rotational_influence}
\end{figure}

Many different types of flow are possible within this parameter space.  Three illustrative examples can be seen in Figure \ref{fig:increasing_rotational_influence}, which shows the specific entropy $s$ for (top row) a non-rotating case at $\mathrm{Ra_F} = 10^4 \mathrm{Ra}_c$ with a moderate density stratification ($N_\rho = 1.4$),  (middle row) a rotating case ($Ta=10^8$) with the same stratification but $Ra_F=56.2 Ra_c \approx 3.8 \times 10^{7}$, and (bottom row) a rotating case ($Ta=10^8$) at $Ra_F = 10^4 Ra_c \approx 6.8 \times 10^{9}.$ 

In our non-rotating simulations, for which a typical case is shown in the topmost panel of Figure \ref{fig:increasing_rotational_influence}, the convection tends to consist of a small number of convective cells, and to be steady in time.  These simple flow patterns persist to surprisingly high values of $Ra_F$ in the setup investigated here (as also seen in Boussinesq simulations with stress-free boundaries in, e.g., \citealt{wang_etal2020}); in some other problem formulations (e.g., with fixed entropy or temperature boundary conditions) the flow tends to become visibly turbulent at lower values of $Ra$ \citep[see examples in, e.g.,][]{anders_brown2017, Rogers_et_al_2003}.  When rotation is dynamically significant, as shown in the middle panel, the convective patterns tend to align with the axis of rotation in accordance with the Taylor-Proudman theorem.  Rotating cases at the same $Ta$ but even higher $Ra_F$ (as shown in the bottom panel), in which the Coriolis force is small relative to inertia,  exhibit time-dependent flow with structure on many spatial scales. 

Most of our simulations behave, qualitatively, like one of the three examples in Figure \ref{fig:increasing_rotational_influence}.  The non-rotating simulations (as sampled in the top panel) represent one extreme; the rotating, very high-$Ra_F$ cases (as in the bottom panel) are another.  The single-celled case is presumably not realised in any actual star, but serves as a useful limit, showing what can occur when a convective plume travels almost unimpeded from the top to the bottom of the domain; in this limit (as we demonstrate below) most of the dissipation occurs in the bottom boundary layer.  The cases with rotation are more realistic, exhibiting flow and dissipation throughout the domain.  Below, we explore (for several different stratifications) how the dissipation and dynamics vary in between these extremes, as a function of rotational influence.  

\section{The magnitude and spatial distribution of viscous dissipation}

\subsection{The maximum value of viscous dissipation at high $Ra_F$}

Here, we examine whether the high levels of dissipation found in \citetalias{CurrieBrowning17} are realised in rotating cases as well. We find that, for the levels of stratification examined here, the total amount of dissipative heating in the rotating simulations appears to approach a similar upper bound to that realised in non-rotating calculations.  The models here were conducted with a different aspect ratio than in \citetalias{CurrieBrowning17} (here the horizontal layer size is twice its depth, whereas in \citetalias{CurrieBrowning17} they were equal) and different boundary conditions (here periodic, impermeable in \citetalias{CurrieBrowning17}), so we also indirectly show that these results are, for an ideal gas equation of state, not directly dependent on these factors.

\begin{figure}
    \centerline{\includegraphics{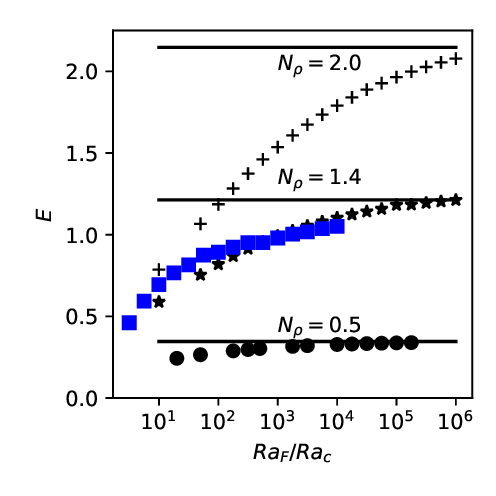}}
    \caption{Calculated values of $E$ at a range of stratifications and supercriticalities for sample non-rotating (black) and rotating cases (blue). Values of $N_\rho = 0.5$ (circles), $1.4$ (stars, and blue squares), and $2.0$ (pluses) are used. The horizontal black lines represent the value of equation (\ref{eq:C&B_lim}) for each value of $N_\rho$. The rotating cases are at latitude $90^{\circ}$ with $Ta=10^8$.  Note that due to the effect of rotation on the critical Rayleigh number, the rotating cases have considerably larger values of $Ra_F$ for a given stratification. }
    \label{fig:E_vs_Ra/Ra_c}
\end{figure}

Figure \ref{fig:E_vs_Ra/Ra_c} shows $E$ for a representative selection of cases at different $Ra_F$ and $Ta$, for three different stratifications. For our non-dimensional setup, $E$ is given by
\begin{equation}
    E = \frac{Pr^2\theta}{ARa_F}\int_V \tau_{ij}\frac{\partial u_i}{\partial x_j} \, dV.
\end{equation}
Recall that $A$ is now non-dimensional, and so for our 2D simulations it is equal to the aspect ratio of the layer, while for the 3D cases $A$ is equal to the aspect ratio squared. The horizontal lines in Figure \ref{fig:E_vs_Ra/Ra_c} show the value of equation (\ref{eq:C&B_lim}) at each value of $N_{\rho}$. At high enough supercriticalities, both the rotating and non-rotating cases appear to approach this limiting value, which is dependent on the layer depth and stratification but independent of $Ra_F$ (and likewise also independent of viscosity or diffusivity).  We have found no cases that exceed this value, but (because it is only an empirical bound) cannot rule out the possibility that it would be exceeded at higher $Ra_F$ or for other parameter regimes.  We have displayed example cases at $Ta=10^{8}$; our cases at $Ta=10^{11}$ and latitude $45^{\circ}$ exhibit identical behaviour.  Both the rotating and non-rotating cases shown here exceed $E=1$ (i.e., the total integrated dissipative heating exceeds the imposed luminosity) at high enough $Ra_F$. However, we cannot rule out different asymptotic values of $E$ for the rotating and non-rotating cases, as discussed in more detail below. 

It is clear from Figure \ref{fig:E_vs_Ra/Ra_c} that the empirical bound on dissipative heating given by equation (\ref{eq:C&B_lim}) is approached only for sufficiently high $Ra_F/Ra_c$, and that the value of $Ra_F/Ra_c$ needed to reach the upper bound is different for each $N_{\rho}$. The largest $N_\rho$ cases have not quite reached the asymptotic upper limit as they have not been performed at a high enough supercriticality.

\begin{figure}
\centerline{\includegraphics{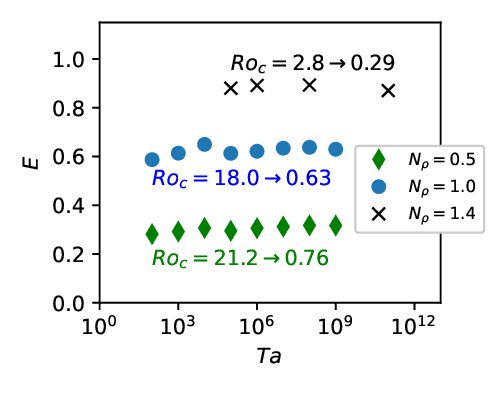}}
    \caption{The values of E for a range of rotation rates. All cases have a fixed supercriticality of $Ra_F/Ra_c = 10^2$; we consider stratifications of $N_\rho = 0.5$, $1$, and $1.4$. The labels show the range of the convective Rossby number $Ro_c$ for increasing values of Ta at each stratification. All simulations shown are at latitude $90^{\circ},$ except for the point at $Ta=10^{11}$ which is at $45^{\circ}$.}
    \label{fig:E_vs_Ta}
\end{figure}

A complementary view but instead focused on the influence of rotation is provided by Figure \ref{fig:E_vs_Ta}, which shows $E$ for a selection of cases at fixed supercriticality (here $Ra_F =10^2 Ra_c$) and three different $N_{\rho}$ but varying $Ta$ (i.e., with varying rotational influence relative to viscous effects). In the regime probed here, it is clear that the presence of rotation does not greatly alter the volume-integrated magnitude of viscous dissipation despite significant changes in the dynamics.  

Note that because rotation stabilises the system against convection, the high-$Ta$ cases shown here have appreciably higher $Ra_F$ than non-rotating cases at the same supercriticality.  We have that $\mathrm{Ra}_c \propto \mathrm{Ta}^{\frac{2}{3}}$ \citep{Chandrasekhar67}, so for example cases at $Ta=10^{11}$ require $Ra_F$ about $10^{7}$ times higher than non-rotating equivalents to reach the same supercriticality. The convective flow fields in the rotating cases are, at equivalent $Ra_F/Ra_c$, more complex than in the non-rotating cases, but they eventually asymptote to similar levels of viscous dissipation. 
Further, the cases shown here span a range of convective Rossby numbers, from $Ro_c \approx 22$ in the lowest-$Ta$ cases to $\approx 0.3$ at $Ta=10^{11}$, and so sample both cases in which rotation plays little dynamical role (with $Ro_c \gg 1$) and those in which it is more significant ($Ro_c < 1$).  

\subsection{Entropy generation by dissipation and conduction}
In the previous section, we suggested that in both rotating and non-rotating cases, the total viscous dissipation at first increases with increased buoyancy driving (higher $Ra_F$) and then plateaus at or below a fixed value (\ref{eq:C&B_lim}) that depends on the layer height and stratification but is independent of the rotation rate or diffusivities.  Here we begin to explore how this arises.  To do so, we consider entropy generation by conduction and dissipation at varying $Ra_F$.  

In a steady state, the energy entering the convection zone at the bottom boundary (by conduction) must equal the energy leaving at the top boundary (also by conduction).  The top boundary is at a lower temperature than the bottom one, so the conductive entropy flux out the top is larger than the entropy flux entering the domain; the difference must be made up by entropy generation within the domain, associated with either conduction or viscous dissipation.  For our simulations (employing entropy diffusion and without magnetic fields), this implies that 

\begin{equation}
     0 = \int_{V} \left[ \nabla \cdot \left(\frac{\bar{\rho} \bar{T} \nabla s }{\bar{T}} \right)  + \frac{ \bar{\rho} \bar{T} \nabla s \cdot \nabla \bar{T}}{\bar{T}^2} + \frac{Pr^2\theta}{Ra_F}\frac{1}{\bar{T}} \tau_{ij} \frac{\partial u_i}{\partial x_j}  \right] dV ,
 \end{equation}
and so
\begin{equation}
    \underbrace{A( e^{N_\rho/m} -1)}_{dS_{out-in})}  = \underbrace{\int_V \frac{\bar{\rho} \bar{T} \nabla s \cdot \nabla \bar{T}}{\bar{T}^2} dV }_{dS_{cond}} + \underbrace{\frac{Pr^2\theta}{Ra_F}\int_V \frac{1}{\bar{T}} \tau_{ij} \frac{\partial u_i}{\partial x_j} dV}_{dS_{diss}},
 \label{eq:cond_entropy_balance}
 \end{equation}
 which follows from equation (\ref{eq:nd_entropy_eq}) after integration (using the divergence theorem and the constraint of mass conservation). We have also used $\left[ -\frac{\bar\rho\bar T \frac{\partial s}{\partial z }}{\bar T}\right]_{z=0}^{z=1} =  e^{N_\rho/m} -1$. For reference, we note that the dimensional equivalent of equation (\ref{eq:cond_entropy_balance}), retaining entropy diffusion, would be 
\begin{equation}
    L\left(\frac{1}{T_{\text{top}}} - \frac{1}{T_0} \right) = \int_V \frac{\kappa \bar{\rho} \bar{T} \nabla s \cdot \nabla \bar{T}}{\bar{T}^2} dV  + \int_V \frac{1}{\bar{T}} \tau_{ij} \frac{\partial u_i}{\partial x_j} dV,
\label{eq:dimcond_entropy_balance}
\end{equation}
and the dimensional equivalent for temperature diffusion, and including magnetism and internal heating, is equation (\ref{eq:entropybound}). 

\begin{figure}
    \centerline{\includegraphics[scale=0.92]{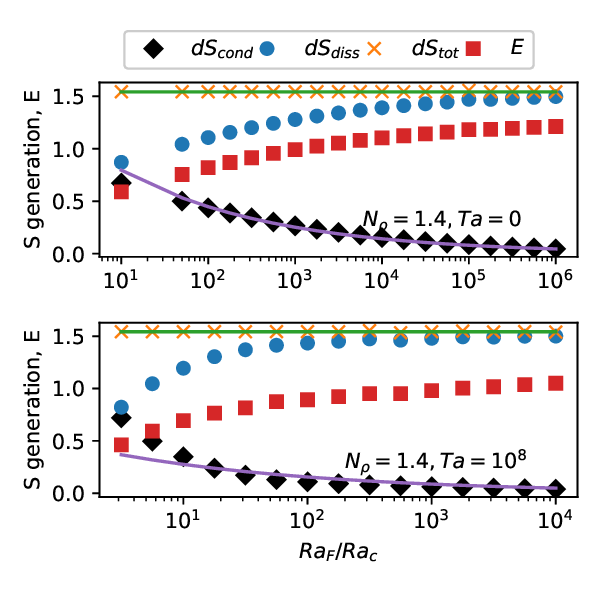}}
    \caption{Entropy generation terms as given in equation (\ref{eq:cond_entropy_balance}) for example non-rotating  (top panel) and rotating (bottom panel) simulations at  $N_{\rho}=1.4$ and varying $Ra_F$.  Shown for each case are the entropy production due to conduction ($dS_{cond}$) and viscous dissipation ($dS_{diss}$), together with their sum ($dS_{tot}$).  Also shown (as horizontal green line) is the entropy mismatch $dS_{out-in}$ for each case; in all cases the sum of $dS_{cond}$ and $dS_{diss}$ matches $dS_{out-in}$.  For comparison, we also plot the value of $E$ for each case, and indicative $Ra_F^{-1/4}$ scaling over the $dS_{cond}$ values; see text for discussion of this scaling.}
    \label{fig:entropygeneration_multicase}
\end{figure}

In Figure \ref{fig:entropygeneration_multicase}, we examine the terms in (\ref{eq:cond_entropy_balance}) for a series of calculations at varying $Ra_F$, $N_{\rho}$, and $Ta$.  In all cases the sum of the terms on the right hand side of equation (\ref{eq:cond_entropy_balance}) ($dS_{cond}$ and $dS_{diss}$) correctly matches $dS_{out-in}$, the mismatch between the entropy flux at the top and bottom boundaries.  As $Ra_F$ increases, the relative contributions of conduction ($dS_{cond}$) and dissipation ($dS_{diss}$) change: at low $Ra_F$ both processes contribute to the entropy balance, whereas at high enough $Ra_F$ there is negligible entropy generation by conduction within the bulk.  

In the non-rotating cases the bulk becomes nearly isentropic at high $Ra_F$, so that the conductive entropy generation term $dS_{cond}$ is then confined mainly to thin thermal boundary layers whose width (discussed in \S \ref{sec:boundarylayerdissipation}) decreases with $Ra_F$.  Thus $dS_{cond}$ scales roughly with the width of these boundary layers; as shown in \S \ref{sec:boundarylayerdissipation}, the largest (top) boundary layer width scales as $Ra_F^{-1/4}$ in our simulations.  We have therefore plotted a corresponding $Ra_F^{-1/4}$ dependence in Figure \ref{fig:entropygeneration_multicase} to guide the eye; in the non-rotating cases $dS_{cond}$ appears to follow this trend reasonably well.  (The line is not a fit; it is chosen to pass through the fourth data point for illustrative purposes.)  The behaviour in the rotating cases is more complicated, as discussed below, partly because in these cases the entropy gradient (and hence also $dS_{cond}$) is nonzero in the bulk.

These trends are linked to the values of $E$ explored above.  We have overplotted the measured values of $E$ at each $Ra_F$ in Figure \ref{fig:entropygeneration_multicase}.  The simulations with the highest $E$ values are those in which entropy generation by conduction is negligible; further, in non-rotating cases the $Ra_F$-dependence of $E$ is well-matched by the $Ra_F$-dependence of $dS_{diss}$ (which in turn is linked to the width of the thermal boundary layers as noted above). 

These results also help us understand why simulations must be run at much higher $Ra_F$ to reach the ``dissipative asymptote" when the stratification is strong (i.e., at high $N_{\rho}$).  The purely conductive state has 
\begin{equation}
dS_{cond} = dS_{out-in} = A(e^{N_\rho/m}  -1),
\end{equation}
which increases with increasing $N_{\rho}$.  

However, knowledge of $dS_{diss}$ alone is not sufficient to determine the actual \emph{value} of $E$ at each $Ra_F$.  In the limit of high $Ra_F$, when $dS_{cond}$ is negligible, we know that
\begin{equation}
    dS_{diss} = \frac{Pr^2\theta}{Ra_F}\int_V \frac{1}{\bar{T}} \tau_{ij} \frac{\partial u_i}{\partial x_j} dV = dS_{out-in} =  A(e^{N_\rho/m}  -1).
\end{equation}
Meanwhile, recall that $E = \frac{Pr^2\theta}{Ra_F A}\int_V \tau_{ij} \frac{\partial u_i}{\partial x_j} dV$, so the highest possible $E$ value consistent with the known $dS_{diss}$ occurs if all the dissipation is at the highest possible temperature.  As noted in \S1, the firm upper bound of \citetalias{Hewitt75} corresponds to this limit.  More generally, the value of $E$ actually attained depends on both the \emph{magnitude} of the dissipative entropy generation term $dS_{diss}$ and on \emph{where it occurs.} For example if $-\tau_{ij}\partial u_i/\partial x_j \equiv Q_0$ were constant throughout the domain, we would have $E = -\frac{Pr^2\theta}{Ra_F}Q_0$, with $dS_{diss} = -\frac{Pr^2\theta}{Ra_F}Q_0 A \int_{0}^{1}(dz/\bar{T})$, which (upon substituting for $\bar{T}$ and integrating) gives
\begin{equation}
    dS_{diss} = -\frac{Pr^2}{Ra_F}\frac{Q_0 A N_\rho}{m}.
\end{equation}
Equating this to $dS_{out-in}$ allows us to solve for $Q_0$ in this limit.  This in turn allows calculation of $E$ for this situation,  
\begin{equation}
    E =  \frac{m\theta}{N_\rho}(e^{N_\rho/m}  -1),
    \label{eq:E_constQlimit}
\end{equation}
which reduces to $E \approx \theta = 1-e^{-N_\rho/m}$, if $N_\rho$ is small. 

Our rotating cases at very high $Ra_F$, which have $Ro_c>1$ and  exhibit intricate flow fields, exhibit $E$ close to the value predicted by equation (\ref{eq:E_constQlimit}), though they  slightly exceed it at the highest $Ra_F$ we have probed. They always remain below the limit of equation (\ref{eq:C&B_lim}).  For example, at $N_{\rho}=1.4$, equation (\ref{eq:E_constQlimit}) yields $E \approx 1.00$, whereas equation (\ref{eq:C&B_lim}) gives $E \approx 1.21$ and equation (\ref{eq:HMWbound}) yields $E \approx 1.54$; our highest-$Ra_F$, $Ta = 10^8$ simulation at that stratification, shown in Figure \ref{fig:E_vs_Ra/Ra_c}, has $E\approx 1.05$. In comparison, many of our non-rotating simulations (which have simpler flow fields) exhibit values of $E$ that exceed equation (\ref{eq:E_constQlimit}), the predicted value of $E$ for uniform dissipation.  For example, at $N_{\rho}=2$, equation (\ref{eq:E_constQlimit}) would yield $E \approx 1.54$, whereas our highest-$Ra_F$ non-rotating simulations at that stratification have $E\approx 2.08$.  This is closer to, but does not exceed, the limit described by equation (\ref{eq:C&B_lim}), which for the same stratification is 2.15. (We have found no cases that exceed the empirical bound in eqn. \ref{eq:C&B_lim}, which is always tighter than the firm bound of eqn. \ref{eq:HMWbound}. For example, at $N_{\rho}=2$, the latter is 2.8, which is significantly larger than in our highest-$Ra_F$ cases.)

If the dissipation were uniformly distributed throughout the domain, equation (\ref{eq:E_constQlimit}) would provide a useful bound on $E$. Some of our simulations exceed this bound, so evidently in at least these cases the dissipation is \emph{not} uniform: a disproportionate amount must occur at higher temperatures, allowing $E$ to be higher than suggested by equation (\ref{eq:E_constQlimit}) while still satisfying the entropy constraint that $dS_{diss} = dS_{out-in}$. In the following section we explore how and when this occurs.

\subsection{The spatial distribution of dissipation}
\label{sec:spatial_distribution_of_diss}

 Here, we determine where the dissipation occurs in our simulations.  We show --- in particular by examination of a ``dissipation half-height" (defined to be the height by which half of the dissipation occurs) --- that the non-rotating cases at high $Ra_F$ which approach the \citetalias{CurrieBrowning17} upper bound correspond to situations in which much of the dissipation occurs close to the bottom of the domain and there is negligible entropy generation by conduction in the bulk.  In rotating cases, the dissipation is more uniformly distributed throughout the interior. 

We begin by defining $L_{diss}(z) = A\int_0^{z}{Q_{diss}dz'}$, where $Q_{diss}$ is the horizontal average of $-\frac{Pr^2\theta}{Ra_F}\tau_{ij} \partial u_i/\partial x_j$ (the local dissipative heating). Here $L_{diss}$ represents the total dissipative heating up to height $z$, so if the heating were uniformly distributed throughout the interior (with $Q_{diss}$ a constant) $L_{diss}/L$ would be a linear function of height, increasing from zero at the lower boundary to $-E$ at the top. Qualitatively, we find that $L_{diss}(z)$ is close to linear for our rotating cases; the dissipative heating is nearly uniform throughout the domain.  In our non-rotating cases, by contrast, much of the dissipation occurs near the lower boundary.  Examples, and a discussion of how these are linked to the buoyancy work and to the dynamics, can be found in \S\ref{sec:energy_balance_and_transport} below.

\begin{figure}
    \centerline{\includegraphics{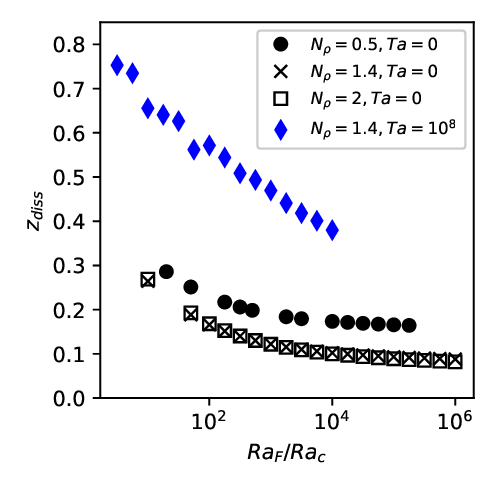}}
    \caption{Values of $z_{diss}$ for a range of non-rotating and rotating simulations at a range of stratifications, $N_\rho$. All rotating cases shown are at latitude $90^{\circ}$.}
    \label{fig:z_diss_half_height}
\end{figure}

A simple, quantitative assessment of the sites where dissipation occurs is provided by Figure \ref{fig:z_diss_half_height}, which shows what we call the  ``dissipation half-height" $z_{diss}$ in a range of cases.  We define this as the location at which $L_{diss}$ reaches \emph{half} its maximum (absolute) value.  That is,
\begin{equation}
    \frac{|L_{diss}(z_{diss})|}{L} =  \int_0^{z_{diss}}{Q_{diss}dz} = \frac{E}{2}.
    \label{eq:z_diss_def}
\end{equation}
If the convection and dissipation were uniform throughout the domain, $z_{diss}$ would be 0.5; meanwhile if the dissipation occurs predominantly at the lower boundary, $z_{diss}$ will tend towards the width of the lower dynamical boundary layer. In both the rotating and non-rotating cases, $z_{diss}$ declines at first with increasing supercriticality: more and more of the dissipation occurs close to the bottom boundary.  In the non-rotating cases it appears to level out (i.e., is approximately constant) at high enough $Ra_F$. In the non-rotating cases studied here --- that is, in 2D and at $Pr=1$ specifically --- at high enough $Ra_F$ the flow consists of a steady cell of convection.  Thus there is dissipation around the single downflow plume, and in the top and bottom dynamical boundary layers, but very little elsewhere in the bulk.  In this limit, $z_{diss}$ is related to the point at which the flow is bent from the vertical towards the horizontal, and this occurs progressively nearer the lower boundary at moderate $Ra_F$.  

In the rotating cases, the dissipation is more uniformly distributed throughout the domain, so $z_{diss}$ is (at any given $Ra_F$) higher than in the non-rotating cases. However, $z_{diss}$ still declines as $Ra_F$ increases.  Note that at fixed $Ta$, as sampled here, increasing $Ra_F/Ra_c$ implies \emph{decreasing} rotational influence on the dynamics, so that the cases at high $Ra_F$ and lower $z_{diss}$ have a higher Rossby number.

\begin{figure}
\centerline{\includegraphics{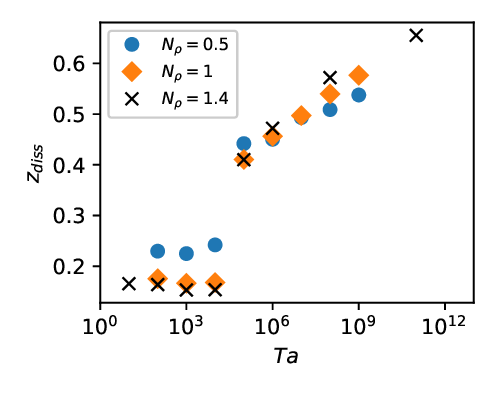}}
    \caption{Values of $z_{diss}$ for a range of simulations at fixed supercriticality, $Ra_F = 10^2 Ra_c$ for stratifications of $N_\rho  = 0.5$, $1.0$, and $1.4$ and a range of $\mathrm{Ta}.$ All cases shown are at latitude $90^{\circ}$, except for the point at $Ta=10^{11}$ which is at $45^{\circ}$.}
    \label{fig:z_diss_fixed_supercrit_increasing_Ta}
\end{figure}

A complementary view is provided by Figure \ref{fig:z_diss_fixed_supercrit_increasing_Ta}, which examines the value of $z_{diss}$ in a series of cases at the same supercriticality $Ra_F =10^2 Ra_c$ but varying $Ta$.  Here we find that $z_{diss}$ increases with increasing $Ta$ (i.e., with increasing rotation rate).  That is,  stronger rotation leads to more of the dissipation occurring in the bulk of the domain, far from the lower boundary. There is a fairly sharp transition between a ``low-$z_{diss}$" state at low $Ta$ (high Rossby number) to a ``high-$z_{diss}$" state for higher $Ta$; beyond this, $z_{diss}$ increases slowly with increasing $Ta$ (i.e., with increasing rotational influence).  This transition is connected to the transition from single-cell states (as achieved in non-rotating cases or at very low $Ta$) to much more intricate, time-dependent flows realised at higher $Ta$ and $Ra_F$.  

These changes in the spatial distribution of the dissipation must be linked to changes in the flow field.   Since the local dissipative heating is related to the stress tensor $\tau_{ij} \partial u_i/\partial x_j$, changes in either the magnitude of $u$, or in the characteristic lengthscales present in the flow, will affect $Q_{diss}$.  Both these quantities are expected to depend on rotation rate \citep[e.g.,][]{Aurnou_et_al_2020, Currie_et_al_2020, nicoski_connor_calkins2023}, so it is natural that the dissipation exhibits some dependence on this as well.  However, $L_{diss}$ must obey the bounds described in \S4.2 at all rotation rates.

Flows in 3D, or in real stars, are bound to exhibit more complexity at all rotation rates, and so we do not expect the numerical values of $z_{diss}$ (for example) to be the same in such cases. Equivalently, the relative proportion of bulk and boundary dissipation might well be different.  (In the context of stellar convection, with no impermeable boundaries, the equivalent of "boundary" dissipation might be convective plumes that are dissipated only when they reach adjacent stably-stratified layers.) The extremely low-$z_{diss}$ states seen here at low $Ta$ are also unlikely to be realised in any real star, since they occur only for single-cell flows with little bulk dissipation.   However, we expect that both the thermodynamic bounds discussed here, and the general trend towards increasing dissipation in the bulk at higher rotation rates, may be robust.  

\section{Links between dynamics, heat transport, and dissipation}

In a steady state, dissipation and dynamics are linked, so insight into either one yields constraints on the other.  Here, we briefly explore how systematic variations in the governing parameters of this problem (namely $Ra_F$, $Ta$, and $N_{\rho}$) lead to changes in the energy transport and in the flow fields, and we explore how these are related to changes in the magnitude and spatial distribution of the dissipation. Our discussion here is also intended to help place our work in context with a large body of previous research on heat transport in both non-rotating and rotating convection.  

In stellar astrophysics, the main purpose of a convective theory is to
provide estimates of the entropy gradient needed to carry a certain
luminosity outwards (e.g., \citealt{Gough_Weiss_1976}). For example, in
standard stellar evolution theory, the radius of a star depends on its
specific entropy, and how this varies with depth (see, e.g., discussions in
\citealt{Ireland_Browning_2018}). There is also substantial astrophysical
interest in properties of the flow itself --- e.g., its magnitude at each
depth --- since these in turn will affect mixing, the transport of heat and
angular momentum, and the generation of magnetic fields.  Hence, we focus
our discussions here on the heat transport, on the related question of how
entropy varies with height in our simulations, and on the magnitude of the
flows themselves.  In Section \ref{sec:predictingKEflux},  we demonstrate that a quantity of particular interest, the kinetic energy flux, can be estimated given knowledge of the dissipative heating.

\subsection{Energy balances and transport terms}
\label{sec:energy_balance_and_transport}

In this section we begin to quantify the links between energy transport in our simulations, and where the dissipation occurs. 

\begin{figure}
\centerline{\includegraphics[width=\columnwidth]{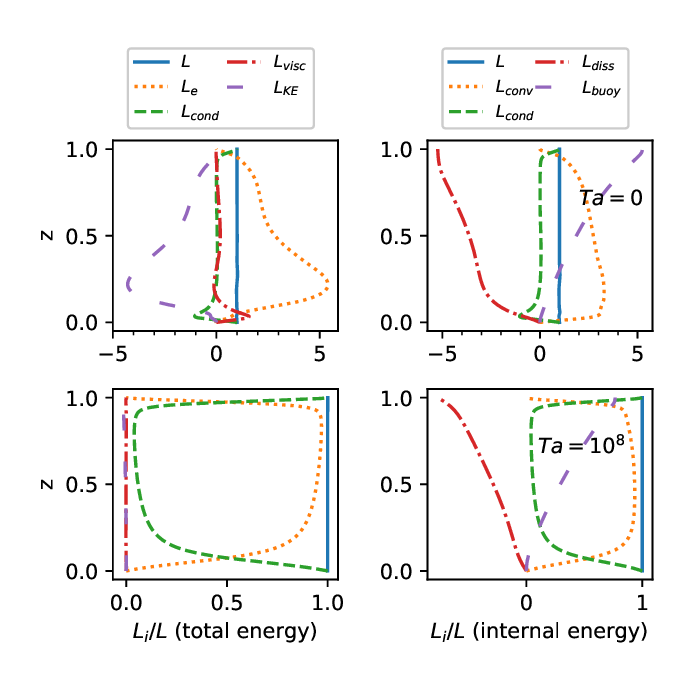}}
    \caption{Fluxes of energy provided by different transport terms in the total energy equation (left panels) and internal energy equation (right panels), for example 2D cases at (top) $N_{\rho}=4$ without rotation and (bottom) $N_{\rho}=1.4$ with rotation ($Ta=10^{8}$). }
    \label{fig:fluxbalance}
\end{figure}

One view of this is provided by Figure \ref{fig:fluxbalance}, which assesses the energy 
transport in two example calculations. The top panels consider a non-rotating case at $N_{\rho}=4, Ra_F=10^{5} Ra_c$; the bottom ones are for a rotating case with $N_{\rho}=1.4, Ra_F = 17.8 Ra_c, Ta=10^{8}$.  

We assess the transport in two complementary ways.  The right panels show the terms arising in the following equation, which arises after integration and manipulation of the total energy equation (eqn. \ref{eq:int_E_eq}): 
\begin{align}L=&FA=\underbrace{Pr\int_{S_{z}}\bar\rho\bar{T}sw\,dS}_\text{$L_{conv}=AF_{conv}$}+\underbrace{\int_{S_{z}}-\bar\rho\bar{T}\frac{\partial{s}}{\partial{z}}\,dS}_\text{$L_{cond}=AF_{cond}$}\nonumber\\&+\underbrace{Pr\int_{V_{z}}-s\bar\rho(\mathbf{u}\cdot\nabla)\bar{T}\,dV}_\text{$L_{buoy}=A\int_0^{z}{Q_{buoy}dz'}$}+\underbrace{\frac{Pr^2\theta}{Ra_F}\int_{V_{z}}-\tau_{ij}\frac{\partial{u_i}}{\partial{x_j}}\,dV}_\text{$L_{diss}=A\int_0^{z}{Q_{diss} dz'}$},\label{Feq}\end{align}
where the volume integrals are over the volume enclosed between $z'=0$ and $z'=z$ and the surface integrals are over the bounding surfaces of that volume. The
convective heat flux  ($F_{conv}$) is defined by the first term; the
conductive heat flux ($F_{cond}$) by the second; the third and fourth terms
define heating and cooling terms ($Q_{diss}$ and $Q_{buoy}$) arising from
the viscous dissipation and from work done against the background
stratification, respectively (as also discussed in \S \ref{sec:spatial_distribution_of_diss}). As noted in \citetalias{CurrieBrowning17}, these latter two terms must balance
when integrated over the entire layer, but they do not have to balance at
each depth.  

The left panels instead arise from considering the total energy balance (i.e.,
including kinetic energy as well as internal), which in a steady state may be written  as (see e.g., \citealt{Viallet13}) 
\begin{align}L=&FA=\underbrace{\int_{S_{z}}Pr\bar\rho sw + \frac{Pr^2\theta}{Ra_F} \bar\rho 
 w \tilde{p} \,dS}_\text{$L_e=AF_{e}$}+\underbrace{\int_{S_{z}}-\bar\rho\bar{T}\frac{\partial{s}}{\partial{z}}\,dS}_\text{$L_{cond}=AF_{cond}$}\nonumber\\&+\underbrace{\frac{Pr^2\theta}{Ra_F}\int_{S_{z}}\frac{1}{2}\bar\rho|u^2|w\,dS}_\text{$L_{KE}=AF_{KE}$}+\underbrace{\frac{Pr^2\theta}{Ra_F}\int_{S_{z}}-(\tau_{ij}{u_i})\cdot{\mathbf{\hat{e}_z}}\,dS}_\text{$L_{visc}=AF_{visc}$},\label{FHeq}\end{align}
defining the enthalpy flux ($F_e$), the kinetic energy flux ($F_{KE}$) and
the viscous flux ($F_{visc}$).  It is common for global-scale simulations
of stellar convection to decompose the transport in this way (e.g., \citealt{Browning_etal2004}, \citealt{Featherstone_Hindman_2016}).  In the notation here, and in Figure \ref{fig:fluxbalance}, positive fluxes
are defined to be vertically upwards.

Whether considering the total or internal energy equations, in a steady state the sum of the
transport terms must equal $L$, the total luminosity, which is constant throughout the layer. The sum of the
transport terms is indicated in Figure \ref{fig:fluxbalance} by a solid line, which is indeed
constant with depth in all the sampled cases. In general, we use $L(z)$ as a gauge of whether a given simulation has been evolved for a long enough time, and averaged over long enough intervals, for the results to be time-independent. We evolved all cases in this paper long enough for $L(z)$ to vary by less than one percent across the layer, and for other aspects of the dynamics (e.g., the kinetic energy evolution) to equilibrate as well.  This means that the simulations were evolved for typically tens of viscous diffusion times, and averaged over intervals ranging from 0.1 to several diffusion times.

The energy transport differs substantially in our non-rotating and rotating cases. The top row in Figure \ref{fig:fluxbalance} is an example of what can occur in non-rotating, stratified cases: here (left panel) the enthalpy flux exceeds the total flux in magnitude; this excess is compensated largely by a negative (inward-directed) kinetic energy flux.  Broadly similar transport has been observed in simulations of stratified convection for decades (see, e.g. \citealt{Hurlburt_et_al_1984} in 2D; \citealt{Stein_Nordlund_1989}; \citealt{Featherstone_Hindman_2016}). Transport by conduction is small throughout the layer, outside of narrow boundary layers.  

By contrast, in the example rotating case (bottom row) the kinetic energy flux is negligible; the enthalpy flux is approximately equal to the total luminosity, with conductive transport small outside the boundary layers.  At this particular $Ra_F$, the conductive boundary layers are still relatively large, and there is evident asymmetry between the top boundary layer and the bottom one (which have different widths).

The connection between this transport and the viscous dissipation is made clearer by comparison to the right panels of Figure \ref{fig:fluxbalance}, which considers the internal energy decomposition for the same cases.  In all cases the buoyancy work is fairly evenly distributed throughout the domain --- that is, outside of the bottom boundary layer $L_{buoy}$ rises linearly towards the top domain.  The rotating case has fairly uniform dissipative heating, with $L_{diss}$ also nearly linear. But in the non-rotating case, the dissipative heating $L_{diss}$ is non-uniform: more of the dissipation is occurring near the bottom boundary.  At the upper boundary, both $L_{buoy}/L$ and $L_{diss}/L$ must approach $+/- E $, respectively, and they do so in both cases; what differs in the rotating and non-rotating simulations is the spatial distribution of $L_{diss}$.  The convective luminosity $L_{conv}$, as defined and plotted here, is often larger than the total luminosity in the non-rotating cases, in accord with the fact that $L_{diss}$ is greater in magnitude than $L_{buoy}$ throughout much of the convection zone.  In the rotating cases, where there is approximate local balance (as well as an exact global balance) between the dissipative heating and the ``cooling'' by buoyancy work, the convective luminosity is closer to unity.

\subsection{Predicting the kinetic energy flux from the viscous
  dissipation }
  \label{sec:predictingKEflux}

The transport revealed here differs in some important ways from that
envisioned in MLT, and some of these differences are
connected to where the viscous dissipation occurs. In this section we consider
the kinetic energy flux, which is not explicitly included in classical MLT \citep[e.g.,][]{Gough_Weiss_1976} but is a
robust feature of stratified convection in stellar environments.  Generally
we find
that the enthalpy flux exceeds the total luminosity by a considerable
amount, and is compensated for by the inward-directed KE flux.  However, prior work has not clearly established what sets the amplitude of these fluxes.  Might it be possible, for example, for a star like the Sun to have a thousand solar luminosities moving outwards in the enthalpy flux, and 999 moving inwards via the KE flux?  In this section we show that knowledge of the viscous dissipation can answer this question, and more generally provide constraints on the magnitude of the kinetic energy flux.

Following \citetalias{CurrieBrowning17}, we define $F_{other}=\int_0^{z}{(Q_{buoy}+Q_{diss}) dz'}$, so that if
conduction is negligible the
total flux $F\approx{F_{conv}}+F_{other}$. Equivalently, $F_{other}=F_{p}+F_{KE}+F_{visc}$, where
$F_{p}=\frac{1}{A}\int_{S_{z}}wp\,dS$. Hence
$F_{other}$ is equivalent to the steady-state transport associated with
processes other than the convective flux as defined above. Outside of the
boundary layers, prior work has found that $F_{KE}$ is generally larger in magnitude than
$F_p$ or $F_{visc}$ (e.g., \citealt{Viallet13}) , so that in the bulk $F_{other} \approx F_{KE}$.  

If the local dissipation and buoyancy work terms \emph{do} balance at each depth, then
$F_{other}$ is zero. This in turn implies negligible kinetic energy
flux. This is approximately the state attained in some of our rotating
cases: there, both $L_{diss}$ and $L_{buoy}$ are linear in $z$, and of
similar magnitude, so that $F_{other} \ll F$.

In our non-rotating 2D cases, by contrast, the concentration of much of the
dissipative heating near the bottom boundary implies a substantial mismatch
between dissipative heating and buoyancy work throughout much of the bulk, so $F_{other}$ is
no longer negligible.  This in turn requires a substantial
kinetic energy flux.  

We can use these ideas to place simple bounds on the magnitude of the kinetic energy flux. Consider the extreme case in which
\emph{none} of the viscous dissipation occurs in the bulk (i.e., it is all
in a comparatively narrow bottom boundary layer).  At some depth
just \emph{above} this boundary layer, nearly all the integrated
viscous heating will have occurred, but very little of the integrated
work will have; in the notation
employed here, $L_{buoy}/L$ will be close to zero, while $L_{diss}/L$ will be
nearly equal to its value at the top of the domain.  The latter is bounded
by $E = d/\hat{H}_T$, as discussed above, so we have $L_{other}/L \approx
E$ just above the bottom boundary.  Hence, if  the ``other'' transport is dominated by the KE flux (rather
than $L_p$ or $L_{visc}$) we expect the maximum absolute value of $L_{KE}/L$ to be
bounded by the value of $E$ at each stratification.

\begin{figure}
\centerline{\includegraphics{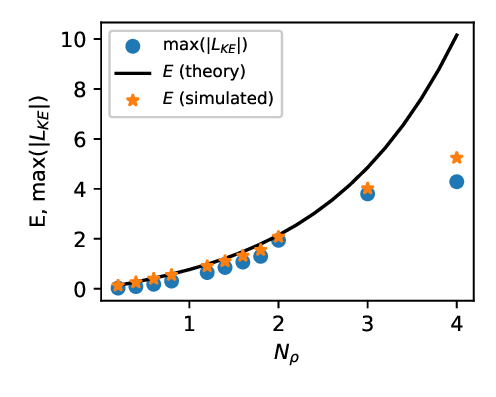}}
  \caption{Comparison between KE flux amplitudes in theory and simulation.  Shown as a solid line is the estimate of eqn. (\ref{eq:C&B_lim}) for $E$ in non-rotating convection at very high $Ra_F$, which we argue in the text provides a bound on the maximum KE flux amplitude.  Also shown are the values of $E$ in sample non-rotating simulations (at high but finite supercriticality) and the maximum absolute values of the KE flux in the same simulations.  }
    \label{fig:KEflux_predictions}
\end{figure}

We examine this prediction in Figure \ref{fig:KEflux_predictions}, which shows the maximum absolute value of the kinetic energy flux in a series of calculations at varying $N_{\rho}$
at high supercriticality ($Ra_F \geq 10^4 Ra_c$), along with the limiting value of $E$
given by equation (\ref{eq:C&B_lim}) and the actual value of $E$ attained in the simulation. For cases at $N_{\rho}$ of two or less, the measured $E$ values adhere closely to the limiting value (indicated by the black line); at the highest $N_{\rho}$, the measured values are lower than the theory. The KE flux closely tracks the measured value of $E$ at each stratification (lying slightly below it), in keeping with the simple model described above.  

These results suggest that at high enough $Ra_F$ the \emph{maximum} possible amplitude of the kinetic energy flux may be estimated simply by calculating $E = d/\hat{H}_T$ (eqn. \ref{eq:C&B_lim}). Our non-rotating cases, which consist of simple uni-cellular flows, actually approach this limit.  However (as noted above) in rotating cases the dissipative heating more nearly balances the buoyancy work at each height, leading to significantly smaller kinetic energy fluxes.  Likewise, more complex flows (as likely realised at higher $Ra_F$ in real stars) likely lead to smaller KE fluxes as well.  We speculate, though, that the limits on the KE flux developed here are unlikely to be \emph{exceeded} by real convective flows. For this to occur, the dissipation and buoyancy work would have to be even more imbalanced than in our single-plume non-rotating cases; for uniform buoyancy work, this would require the dissipation to be concentrated to an even smaller part of the domain than in these simulations.

\subsection{Entropy profiles and Nusselt number scalings}
\label{sec:entropy_profiles_and_nusselt_scalings}

In  previous sections, we saw that the energy transport in our simulations --- and in particular the relative contributions of conduction and convection --- varied in response to changes in the key controlling parameters $N_{\rho}$, $Ra_F$, and $Ta$.  Here we explore how these variations arise, and in particular how they are linked to the entropy gradients established by the convection. 

In our work entropy is fixed at the upper boundary; at the lower boundary its gradient is fixed. Recall also that we have assumed that conductive transport is proportional to entropy gradients (rather than temperature gradients).  Together, these imply that in the \emph{absence} of convection, we would expect a linear specific entropy profile, extending from $s=0$ at the top to $\Delta s_{cond}$ at the bottom. For the models considered here, $\Delta s_{cond}$ is
\begin{equation}
\Delta s_{cond} = \frac{1}{\theta m} [e^{N_{\rho}} -1 ].
\end{equation}

\begin{figure}
\centerline{\includegraphics[scale=0.95]{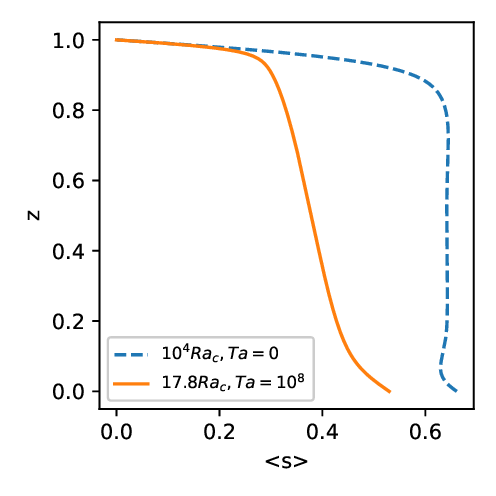}}
    \caption{Horizontally-averaged specific entropy as a function of depth in two example cases, both at $N_{\rho}=1.4$. The solid (orange) line is for a rotating case at $Ta=10^8$ and the dashed (blue) line is for a non-rotating case ($Ta=0$).}
    \label{fig:sprofiles_vsdepth}
\end{figure}

If convection occurs, a smaller total entropy contrast between the top and bottom boundaries, $\Delta s$, is required to carry the same imposed $F$.  In our simulations the total $\Delta s$, and its variation with height, are functions of $N_{\rho}$, $Ra_F$, and $Ta$. We assess these for two illustrative cases in Figure {\ref{fig:sprofiles_vsdepth}}, which shows $\langle s \rangle(z)$ (where $\langle\cdot\rangle$ denotes a horizontal average) for both a rotating case (at $Ta=10^8$, $Ra_F = 17.8 Ra_c$) and a rotating one (with $Ra_F = 10^4 Ra_c$) at $N_{\rho}=1.4$. In both cases, conduction must carry all the energy within some distance of the boundaries, so there are steep entropy gradients at the top and bottom of the domain; the entropy gradient is smaller in the bulk. The total $\Delta s$ across the layer is similar in the two cases, but its spatial variation is different: in the non-rotating the interior is nearly isentropic ($\langle s \rangle$ is close to a vertical line), whereas in the rotating case there is a visible, finite slope to $\langle s \rangle$ throughout the bulk.

To analyse these trends  quantitatively, we turn to an average measure of the heat transport over the domain (rather than to its spatial variation).  In studies of Rayleigh-B\'enard convection, it is customary to encapsulate this via the Nusselt number $Nu$, a dimensionless measure of the  heat transport relative to that provided by conduction.  There is no universally-accepted definition of $Nu$ that makes sense for all boundary conditions, stratifications, and with/without rotation, but sensible definitions have the property that they are large when convection is efficient, and tend to one as the convection vanishes (i.e., as all transport becomes conductive).  For the mixed fixed-flux, fixed-entropy boundary conditions here, we choose to adopt 
\begin{equation}
    Nu = \frac{\Delta s_{cond}}{\Delta s}
\end{equation}
as our definition of $Nu$.  This is a global measure of the efficiency of the convective flow --- more efficient convection should have a smaller $\Delta s$ across the layer, and so a higher $Nu$ --- but normalised to the $\Delta s_{cond}$ that would be required to carry the flux in the absence of convection.  This is akin to the definition for Boussinesq convection adopted by, e.g., \citet{Kazemi_et_al_2022}.  

\begin{figure*}
    \centerline{\includegraphics{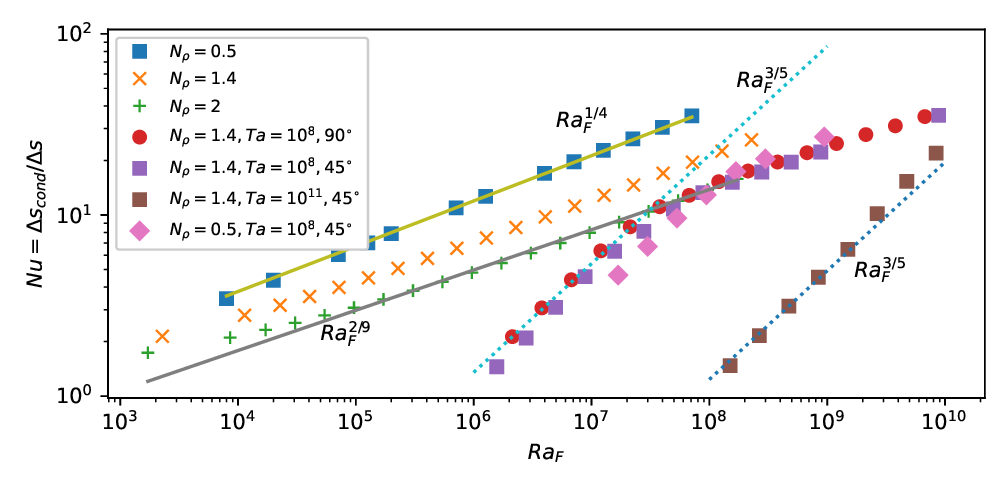}}
    \caption{Dimensionless heat transport (Nusselt number) as a function of $Ra_F$ for a sample of rotating and non-rotating cases at different $N_{\rho}$.}
    \label{fig:Nu_vs_Raf}
\end{figure*}

The resulting measures of $Nu$ are plotted for a sample of cases with varying $Ra_F$, $N_{\rho}$, and $Ta$ in Figure \ref{fig:Nu_vs_Raf}. We have also overplotted several previously-proposed scaling relations, as discussed below.  We have chosen here not to normalise each case by $Ra_c$, primarily because $Ra_c$ varies so much across the simulations sampled here; in general, each ``track" of simulations shown begins with $Ra_F$ of order ten times critical at that $N_{\rho}$ and $Ta$.

First, consider the non-rotating, weakly-stratified cases at $N_{\rho} = 0.5$. These are well-matched by the power law $Nu \propto Ra_F^{1/4}$, which is obtained if transport within the bulk is entirely by convection, transport within narrow thermal boundary layers is by conduction, and the width of the boundary layers is set by the requirement that they be marginally stable against convection \citep{Malkus_1954}. The scaling at higher $N_{\rho}$ appears to be slightly less steep than this.  For comparison, we have overplotted $Nu \propto Ra_F^{2/9}$.  This scaling would arise if $Nu \propto Ra^{2/7}$ (from $Ra_F = NuRa$),  as has often been reported in non-rotating experiments and simulations (see, e.g., discussions in \citealt{Grossmann_Lohse_2000}; \citealt{Siggia_1994}).  None of our data are consistent with the so-called ``ultimate regime" scaling $Nu \propto (RaPr)^{1/2}$ \citep[e.g.][]{chavanne_etal1997}, which has been conjectured to apply at very high $Ra$.

The rotating cases exhibit steeper $Nu(Ra_F)$ scalings. Figure \ref{fig:Nu_vs_Raf} shows a series of cases at fixed $Ta=10^{8}$ and a smaller number of cases at $Ta=10^{11}$.  Note that the cases at $Ta=10^{11}$ are situated at 45$^{\circ}$ rather than 90$^{\circ}$.  At moderate $Ra_F$ (where $Ro_c$ is small) the dependence of $Nu$ on $Ra_F$ appear to be reasonably well described by the overplotted scaling $Nu \propto Ra_F^{3/5}$. At higher $Ra_F$, when rotation is unimportant dynamically (large $Ro_c$), the $Ta=10^{8}$ cases latch on to the non-rotating scalings. The slope of the $Nu(Ra_F)$ relation is similar for the cases at $Ta=10^8$ and $Ta=10^{11}$.  Note that for fixed $Ta$, increasing $Ra_F$ is equivalent to increasing $Ra_F/Ra_c$, so a plot of $Nu$ as a function of $Ra_F/Ra_c$ would exhibit the same slope.  

This behaviour is consistent with prior results in different settings.  For example, the transition from a steep ``rotating" $Nu(Ra)$ relation to a shallower ``non-rotating" one was reported by \citep{king_etal2009} to occur in plane-layer experiments and accompanying Boussinesq simulations (with fixed temperature and no-slip boundaries); see also discussions in \citet{King_et_al_2013} and \citet{Aurnou_et_al_2020}. Simulations in spherical shells (see discussions in \citealt{Gastine_et_al_2016}, \citealt{long_etal2020_scaling}, discussing fixed-temperature and fixed-flux simulations respectively) likewise exhibit similar trends. 

Interestingly, the $Nu \propto Ra_F^{3/5}$ scaling seen in our rotating cases is in accord with the expectations of rotating mixing-length theory (\citealt{Currie_et_al_2020}, \citealt{Barker_et_al_2014}, \citealt{Stevenson_1979}).  The same scaling law also arises in the classical ``CIA balance," which supposes a dynamical balance between Coriolis, inertial, and buoyancy terms in the momentum equation (\citealt{Aurnou_et_al_2020}, \citealt{Vasil_et_al_2021}) and in asymptotic theories of convection at low Rossby number \citep{Julien_et_al_2012}. In these theories, the transport is predicted to follow $Nu \propto (Ra_F/Ra_c)^{3/5} \propto (Ra_F Ta^{-2/3})^{3/5}$, which (upon substituting in the definitions for $Ra_F, Nu$, and $Ta$) is diffusion-free.  This suggests that in the rapidly-rotating limit the diffusive boundary layers are playing a less significant role in the heat transport.  In dimensional terms, this scaling implies that the entropy gradient in the bulk of the convection zone  becomes steeper when rotation is more rapid, scaling as $ds/dz \propto \Omega^{4/5}$, with $\Omega$ the angular velocity.

\subsection{Boundary layers and the link to dissipation}
\label{sec:boundarylayerdissipation}

The trends explored above arise partly from the varying influence of viscous and thermal boundary layers in our simulations.  In this subsection, we explore how the widths and other properties of these boundary layers vary as the supercriticality of the convection, the density stratification, and the rotational influence are changed. We also discuss the manner in which the boundary layers, heat transport, flow amplitudes, and dissipation are linked --- and demonstrate explicitly that knowledge of some of these aspects constrains the others.

Many different definitions of the boundary layers have been employed in the literature on Boussinesq convection, but our inclusion of rotation, our use of a fixed-flux thermal boundary condition at the bottom boundary, and our adoption of stress-free velocity boundary conditions together implies that some of these definitions are not relevant (see discussion in \citealt{Long_et_al_2020}).  We choose here to adopt the simple method suggested by \citet{Long_et_al_2020}, defining the width of these layers (near the top and bottom of the domain) to be the points at which the advective and conductive contributions to the heat transport are equal.  Inside the boundary layer conduction dominates the heat transport; outside it convection does. \citet{Long_et_al_2020} demonstrate that this method gives sensible results in a variety of Boussinesq settings (with and without rotation), though to our knowledge it has not been previously employed to study anelastic convection simulations.  

At the top and bottom boundaries conduction must carry all the energy (because in our simulations the vertical convective velocity goes to zero there, and because near-surface radiative cooling has been ignored). The value of the horizontally averaged entropy gradient $d \langle s \rangle/dz$ at the bottom boundary is therefore determined by the energy flux entering the domain.  At the top boundary, the entropy is fixed (rather than its gradient), but in a steady state the simulation must still develop a sufficiently large entropy gradient to carry the same energy flux out the top boundary. Specifically, because we have assumed that conduction diffuses entropy, we must have
\begin{equation}
    F = F_{cond} = -\bar{\rho} \bar{T} \frac{d \langle s \rangle}{d z}
\end{equation}
at both the top and bottom boundaries. (As elsewhere, all variables here are dimensionless; the dimensional version would have a factor of $\kappa$ on the right hand side of the equation.)  Because we are considering stratified convection, the density at the top of the domain is smaller than at the bottom, so we expect the entropy gradients $d \left< s \right>/dz$ that develop will be greater at the outer boundary than at the inner one.

We now suppose that within the conductive boundary layers $d\langle s \rangle/dz$ is approximately uniform, and equal to 
\begin{equation}
    \frac{d\langle s \rangle}{dz} \approx \frac{\Delta s_{bl}}{\delta_{bl}} \approx -\frac{F}{\bar{\rho} \bar{T}}
\label{eq:dsbl_def}
\end{equation}
where $\Delta s_{bl}$ is the entropy jump across the boundary layer and $\delta_{bl}$ is its width, $\bar{\rho}$ and $\bar{T}$ are evaluated at the top or bottom of the domain (for the top and bottom boundary layers respectively), and where we have assumed conduction carries all the flux within the boundary layer. 

\begin{figure}
    \centerline{\includegraphics{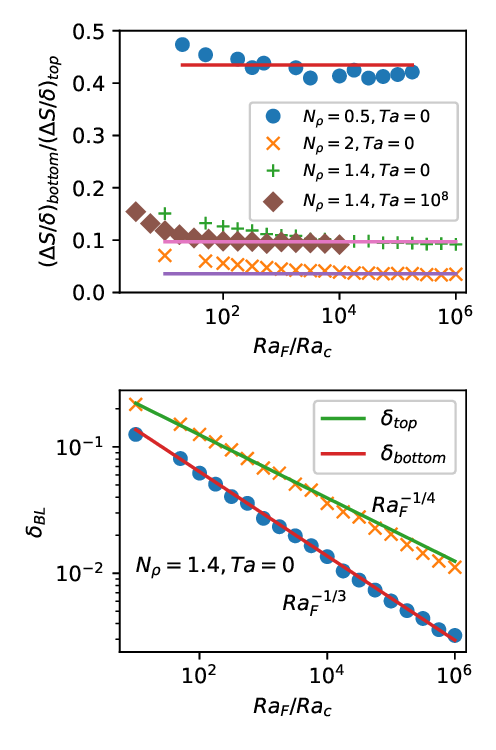}}
    \caption{Properties of top and bottom boundary layers at varying $Ra_F$, $N_{\rho}$, and $Ta$.  Top panel shows the ratios of the dimensionless entropy gradients across the bottom and top boundary layers (as defined in the text); bottom panel shows the width of the top and bottom boundary layers in example non-rotating cases at $N_{\rho}=1.4$. The horizontal lines show the result of evaluating equation (\ref{eq:BLratio}) at each stratification.}
    \label{fig:boundarylayer_widths}
\end{figure}

In Figure \ref{fig:boundarylayer_widths} (top) we compare the resulting predictions for $\Delta s_{bl}/\delta_{bl}$ to measurements in example simulations.  We show the ratio of $\Delta s_{bl}/\delta_{bl}$ in the bottom boundary layer to that in the top; the horizontal lines denote 
\begin{equation}
\label{eq:BLratio}
    \frac{\bar{\rho}_{\mathrm{top}}\bar{T}_{\mathrm{top}}}{\bar{\rho}_{0} \bar{T}_{0}}
\end{equation}
for each stratification, which this ratio should approach (per eqn. \ref{eq:dsbl_def}, and given that $F$ is the same at both boundaries). The agreement between the measured values and the estimated value is good at high $Ra_F$ for all stratifications shown, in both rotating and non-rotating cases.  At low $Ra_F$ the agreement is less good. With increasing $N_{\rho}$ the boundary layers become increasingly asymmetric, so that for example in cases with $N_{\rho}=2$, $\Delta s_{bl}/\delta_{bl}$ is more than 25 times larger in the top boundary layer than in the bottom.  This, again, is a consequence of the much smaller densities and temperatures at the top of these stratified domains, which then require a much larger entropy gradient to carry the imposed flux out the top boundary.

In the bottom panel of Figure \ref{fig:boundarylayer_widths} we show how the top and bottom boundary layer widths vary with $Ra_F$ for example cases at $N_{\rho} = 1.4$. As expected, the boundary layers grow thinner at higher $Ra_F$.  The top and bottom boundary layers appear to follow slightly different trends: we have overplotted $\delta_{bl} \propto Ra_F^{-1/4}$, which appears to match the top boundary layer data well, and $Ra_F^{-1/3}$, which matches the bottom boundary layer well.  

These findings are consistent with, and aid in understanding, our findings for the dynamics and heat transport (i.e., $Nu(Ra_F)$ scalings) in previous sections.  In the non-rotating cases at high $Ra_F$ nearly the entire $\Delta \langle s \rangle$ across the whole domain occurs in the top and bottom boundary layers; hence their width determines the overall Nusselt number for the entire domain.  (In rotating cases, the entropy contrast across the bulk can approach or exceed that in the boundary layers; see discussions in, e.g., \citealt{Barker_et_al_2014}, \citealt{Currie_et_al_2020}.)  The top boundary layer is, in our stratified calculations, probably the more restrictive of these because it is thicker; we expect thus expect that in these calculations the Nusselt number should scale approximately as the height of the layer divided by the width of this boundary layer.  Here, this implies that $Nu$ should scale as $Ra_F^{1/4}$, which is in agreement with many of our findings for the non-rotating cases in Section \ref{sec:entropy_profiles_and_nusselt_scalings} above.  

The steeper heat-transport scalings exhibited by rotating cases are linked to where dissipation occurs. This is because the dissipation and the work done against the background stratification must balance; this balance gives rise to the exact relationship in Boussinesq convection between the $Nu(Ra_F)$ relationship and the viscous dissipation \citep{Shraiman_Siggia_2000}, and to a more complex analogue of this in the anelastic case \citep[as shown recently by][]{Jones_et_al_2022}.  Changes in where the dissipation occurs --- which in turn arise because the convective velocities and lengthscales change in the presence of rotation, as discussed above --- thus also give rise to changes in the heat transport.  

This link between dissipation and transport has led some prior authors to separate the viscous dissipation into ``bulk" and ``boundary" contributions --- see, e.g., \citet{Grossmann_Lohse_2000}, \citet{Jones_et_al_2022}, \citet{Gastine_et_al_2016} --- and to posit that transitions in the heat transport correspond to changes between bulk-dominated or boundary-dominated dissipation (\citealt{Grossmann_Lohse_2000}).  In our simulations, we cannot usefully divide the dissipation in this way; because the boundary layers as defined here become very thin at high $Ra_F/Ra_c$, the dissipation is almost always ``bulk-dominated."  We have argued above that $z_{diss}$ provides, for our setup, a more meaningful distinction between cases where the dissipation is concentrated near the boundaries and those where it is distributed throughout the domain.  We find that cases that fall on the rotating scaling relation $Nu \propto (Ra_F/Ra_c)^{3/5}$ systematically have larger $z_{diss}$ than those which follow the non-rotating heat transport scalings ($Nu \propto Ra_F^{1/4}$).

\subsection{Comparison to fully three-dimensional flows}
\label{sec:3dcomparison}
The simulations presented in the preceding figures were restricted to two spatial dimensions (i.e., assuming axisymmetry in one dimension).  In this section we provide a preliminary view of whether the key quantities assessed in this paper -- namely the overall levels of dissipation (as measured by $E$) and its spatial distribution (as encapsulated by $z_{diss}$ and by the spatial distribution of $L_{diss}$) -- are different in 2D and 3D cases.  

\begin{figure}
\centerline{\includegraphics[scale=0.8]{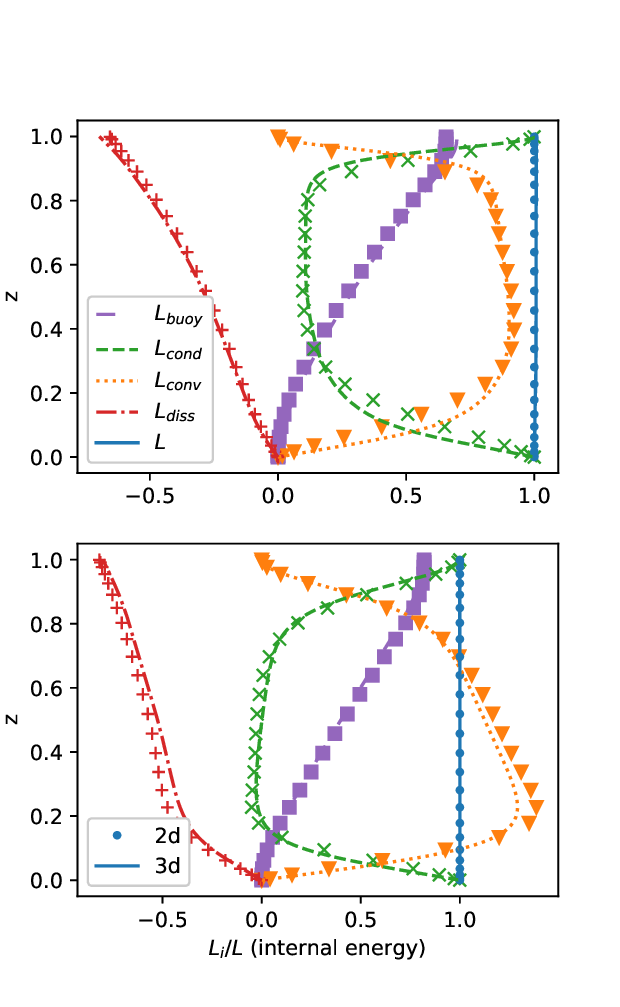}}
  \caption{Internal energy transport in rotating (top) and non-rotating (bottom) cases in both 2D and 3D. The rotating cases are at latitude $45^{\circ}$.}
    \label{fig:3dvs2d}
\end{figure}

In Figure \ref{fig:3dvs2d}, we examine the energy transport in two example 3D cases and their closest 2D equivalents. We display both non-rotating cases (bottom panel, at $Ra_F \approx 2.3 \times 10^{4} \approx 100 Ra_c$) and rotating ones (top panel, at $Ta=10^{6}$, $Ra_F = 3.3 \times 10^{5} \approx 10 Ra_c$, at 45 degrees latitude).  The 2D/2.5D cases have the same $Ra_F$ as the 3D ones.  All cases have $N_{\rho}=1.4$, and an aspect ratio of 2:1 (i.e., the horizontal dimensions extend twice as far as the vertical one).  The 3D cases were run at resolutions of $256 \times 256 \times 128$, and evolved for more than a diffusion time.  In the figure, data from the 2D cases are over-plotted as symbols, with data from the 3D cases as lines.

It is clear that the energy transport in the 3D cases is very similar to that realised in the 2D/2.5D cases.  In the non-rotating case, for example, $L_{diss}$ has a steep slope near the lower boundary, and a smaller one in the bulk;  this reflects the fact that much of the dissipation is occurring near the lower boundary.  In the rotating case, $L_{diss}$ is less sharply peaked, reflecting the more even distribution of dissipation throughout the bulk.  In both cases, the other transport terms ($L_{cond}$, $L_{buoy}$, $L_{conv}$, and the total transport $L$) are also quite similar in 2D and 3D. 

The other bulk quantities of interest to us -- $E$ and $z_{diss}$ -- are also similar in 2D and 3D.  The rotating 3D case shown here has (time-averaged) $E=0.69$ and $z_{diss} = 0.61$; the corresponding 2.5D case had $E=0.66$ and $z_{diss} = 0.60$.  The non-rotating 3D case has $E=0.80$ and $z_{diss} = 0.21$; the corresponding 2D case had $E=0.82$ and $z_{diss} = 0.17$.  Crucially, this suggests that in 3D cases $z_{diss}$ exhibits a similar dependence on rotation as it did in 2.5D: namely, it is significantly larger when rotation is present (because more of the dissipation is distributed in the bulk in that case) than in the non-rotating case (where much of the dissipation is concentrated near the boundaries).  

These results suggest that the key quantities examined in this paper may not be too sensitive to the assumption of axisymmetry.  We defer a more detailed study of the 3D cases to later work.

\section{Discussion and conclusions}

We have presented the first systematic investigation of viscous dissipation in a rotating, stratified plane layer of convection, studied here within the anelastic approximation for an ideal gas.  We have shown that, for fixed convective supercriticality $Ra_F/Ra_c$ and moderate stratification, the total dissipative heating does not depend appreciably on rotation rate.  However, the spatial \emph{distribution} of the dissipation does vary with rotation, and this has a number of important consequences for the dynamics and heat transport. 

The total dissipation is thermodynamically bounded by eqn. (\ref{eq:HMWbound}), which corresponds to the case in which there is negligible entropy generation by conduction in the bulk of the domain and all the dissipation occurs at the highest possible temperature. In practice we have not found any cases in which the dissipation exceeded the tighter, empirical bound of eqn. (\ref{eq:C&B_lim}), which  does not depend directly on the diffusivities or on rotation rate.    Our non-rotating cases, which exhibit  simple mono-cellular flows, approach the latter bound at high enough $Ra_F/Ra_c$.  These represent an extreme case in which very little dissipation occurs in the bulk of the convection zone, so we regard them as a limit on $E$ that is unlikely to be exceeded by more realistic flows.  

In  rotating cases the viscous dissipation is  more uniformly distributed throughout the layer than in corresponding non-rotating cases. In the non-rotating simulations, much of the dissipation occurs near the bottom of the computational domain, so that although there is a global balance between dissipation and work done against the background stratification, these quantities do not balance at each depth.  We defined a new quantity $z_{diss}$, the height at which half the total viscous dissipation has occurred, which encapsulates the spatial distribution of the dissipation in a simple way, and used it to characterise our simulations in different regimes.

We have shown that the heat transport scalings ($Nu(Ra_F)$) in our rotating cases appear to be consistent with theoretical diffusion-free predictions arising from either ``rotating mixing length theory" (\citealt{Stevenson_1979}, \citealt{Barker_et_al_2014}, \citealt{Currie_et_al_2020}) or, equivalently, from a conjectured balance between Coriolis, inertial, and buoyancy forces (e.g., \citealt{Aurnou_et_al_2020}, \citealt{Gastine_et_al_2016}, \citealt{Vasil_et_al_2021}). Prior work has shown this in other settings (mainly within the Boussinesq approximation).

We have shown that these changes in heat transport are linked to where in the domain the dissipation occurs.  This is similar to the case in Boussinesq convection, where prior work (e.g., \citealt{Grossmann_Lohse_2000}) has established that the $Nu(Ra_F)$ heat transport relation varies depending on whether the dissipation is ``bulk" or ``boundary" dominated, and broadly in line with very recent theoretical predictions for the anelastic case \citep{Jones_et_al_2022}.  Here, the situation is more complex than in the Boussinesq case because of the background stratification, but the same basic trends appear to hold.  In particular, we find that cases which follow the ``rotating" heat transport relation are those for which $z_{diss}$ is especially high.

We also established that, for the setup examined here, the thermal boundary layers in our simulations are asymmetric -- the top one being considerably larger than the bottom -- and that the thicknesses of the top and bottom boundary layers scale differently with $Ra_F$.  

Finally, we have explored the link between dissipation and the kinetic energy flux.  We developed a simple model of the kinetic energy flux in our non-rotating cases -- based on the idea that dissipation approaches the upper bound at high enough $Ra_F$, and that much of the dissipation occurs near the lower boundary -- and showed that it provided a reasonably accurate prediction of the maximum (negative) kinetic energy flux attained in our simulations for each stratification at high enough $Ra_F$. We have argued that this provides an upper bound on the kinetic energy flux achievable by real convection.

If our results are applicable to real stars, one conclusion is that rapidly-rotating stars should exhibit less convective overshooting into adjacent stably-stratified regions than slowly-rotating ones.  From the perspective adopted here, this is because in rotating convection zones the buoyancy work at each depth is more nearly balanced \emph{locally} (rather than just globally) by dissipation.  Equivalently, the more rapidly rotating cases have more dissipation per unit volume in the bulk (all else being equal).  For a fixed stratification, this will imply (following the discussion in, e.g., \citealt{Anders_et_al_2022}) that convective motions in rotating stars will have less kinetic energy when they reach the boundary of the Schwarzschild-unstable region, so they will penetrate less deeply into the adjacent layers.  

Similarly, we expect a smaller kinetic energy flux in rotating stars and planets than in non-rotating ones.  The dynamical consequences of this are not yet clear, but we intend to explore them in the future.

\section*{Acknowledgements}

We thank the anonymous referee for a thoughtful and comprehensive review, which significantly improved the paper. SL was supported by a Science and Technology Facilities Council (STFC) PhD studentship (ST/T506084/1). Portions of this work were also supported by the European Research Council under ERC grant agreement no. 337705 (CHASM). LC also gratefully acknowledges support from the STFC (ST/X001083/1). We thank the Isaac Newton Institute for Mathematical Sciences, Cambridge, for support and hospitality during the programme ``Frontiers in Dynamo Theory: From the Earth to the Stars" (DYT2) where work on this paper was undertaken; this was supported by EPSRC grant no EP/R014604/1. We are grateful to numerous participants in this programme for helpful conversations. We are likewise grateful to the Kavli Institute for Theoretical Physics for their hospitality during the early phases of this work; this research was supported in part by the National Science Foundation under Grant No. NSF PHY-1748958. This paper draws on simulations that were carried out on the University of Exeter supercomputer Isca, on the DiRAC Complexity system, and on the DiRAC Data Intensive service at Leicester (DiaL), operated by the University of Leicester IT Services, which forms part of the STFC DiRAC HPC Facility (www.dirac.ac.uk). The DiaL equipment was funded by BEIS capital funding via STFC capital grants ST/K000373/1 and ST/R002363/1 and STFC DiRAC Operations grant ST/R001014/1; the Complexity system was also funded by STFC DiRAC Operations grant ST/K0003259/1. DiRAC is part of the National e-Infrastructure. This work also made use of the Hamilton HPC service of Durham University.  

\section*{Data Availability}

The codes used to produce the simulations in this paper, and selected outputs from the simulations themselves, have been uploaded to the Exeter Open Research repository (ore.exeter.ac.uk) and are available for download there (DOI 10.24378/exe.4945).




\bibliographystyle{mnras}
\bibliography{references} 

\begin{thebibliography}{}
\makeatletter
\relax
\def\mn@urlcharsother{\let\do\@makeother \do\$\do\&\do\#\do\^\do\_\do\%\do\~}
\def\mn@doi{\begingroup\mn@urlcharsother \@ifnextchar [ {\mn@doi@}
  {\mn@doi@[]}}
\def\mn@doi@[#1]#2{\def\@tempa{#1}\ifx\@tempa\@empty \href
  {http://dx.doi.org/#2} {doi:#2}\else \href {http://dx.doi.org/#2} {#1}\fi
  \endgroup}
\def\mn@eprint#1#2{\mn@eprint@#1:#2::\@nil}
\def\mn@eprint@arXiv#1{\href {http://arxiv.org/abs/#1} {{\tt arXiv:#1}}}
\def\mn@eprint@dblp#1{\href {http://dblp.uni-trier.de/rec/bibtex/#1.xml}
  {dblp:#1}}
\def\mn@eprint@#1:#2:#3:#4\@nil{\def\@tempa {#1}\def\@tempb {#2}\def\@tempc
  {#3}\ifx \@tempc \@empty \let \@tempc \@tempb \let \@tempb \@tempa \fi \ifx
  \@tempb \@empty \def\@tempb {arXiv}\fi \@ifundefined
  {mn@eprint@\@tempb}{\@tempb:\@tempc}{\expandafter \expandafter \csname
  mn@eprint@\@tempb\endcsname \expandafter{\@tempc}}}

\bibitem[\protect\citeauthoryear{{Alboussi{\`e}re} \&
  {Ricard}}{{Alboussi{\`e}re} \& {Ricard}}{2013}]{Alboussiere_Ricard_2013}
{Alboussi{\`e}re} T.,  {Ricard} Y.,  2013, \mn@doi [Journal of Fluid Mechanics]
  {10.1017/jfm.2013.241}, \href
  {https://ui.adsabs.harvard.edu/abs/2013JFM...725R...1A} {725, R1}

\bibitem[\protect\citeauthoryear{{Alboussi{\`e}re} \&
  {Ricard}}{{Alboussi{\`e}re} \& {Ricard}}{2014}]{Alboussiere_Ricard_2014}
{Alboussi{\`e}re} T.,  {Ricard} Y.,  2014, \mn@doi [Journal of Fluid Mechanics]
  {10.1017/jfm.2014.336}, \href
  {https://ui.adsabs.harvard.edu/abs/2014JFM...751..749A} {751, 749}

\bibitem[\protect\citeauthoryear{{Alboussi{\`e}re}, {Curbelo}, {Dubuffet},
  {Labrosse}  \& {Ricard}}{{Alboussi{\`e}re}
  et~al.}{2022}]{Alboussiere_et_al_2022}
{Alboussi{\`e}re} T.,  {Curbelo} J.,  {Dubuffet} F.,  {Labrosse} S.,   {Ricard}
  Y.,  2022, \mn@doi [Journal of Fluid Mechanics] {10.1017/jfm.2022.216}, \href
  {https://ui.adsabs.harvard.edu/abs/2022JFM...940A...9A} {940, A9}

\bibitem[\protect\citeauthoryear{{Anders} \& {Brown}}{{Anders} \&
  {Brown}}{2017}]{anders_brown2017}
{Anders} E.~H.,  {Brown} B.~P.,  2017, \mn@doi [Physical Review Fluids]
  {10.1103/PhysRevFluids.2.083501}, \href
  {https://ui.adsabs.harvard.edu/abs/2017PhRvF...2h3501A} {2, 083501}

\bibitem[\protect\citeauthoryear{{Anders}, {Manduca}, {Brown}, {Oishi}  \&
  {Vasil}}{{Anders} et~al.}{2019}]{Anders_et_al_2019}
{Anders} E.~H.,  {Manduca} C.~M.,  {Brown} B.~P.,  {Oishi} J.~S.,   {Vasil}
  G.~M.,  2019, \mn@doi [\apj] {10.3847/1538-4357/aaff61}, \href
  {https://ui.adsabs.harvard.edu/abs/2019ApJ...872..138A} {872, 138}

\bibitem[\protect\citeauthoryear{{Anders}, {Jermyn}, {Lecoanet}  \&
  {Brown}}{{Anders} et~al.}{2022}]{Anders_et_al_2022}
{Anders} E.~H.,  {Jermyn} A.~S.,  {Lecoanet} D.,   {Brown} B.~P.,  2022,
  \mn@doi [\apj] {10.3847/1538-4357/ac408d}, \href
  {https://ui.adsabs.harvard.edu/abs/2022ApJ...926..169A} {926, 169}

\bibitem[\protect\citeauthoryear{{Arnett}, {Meakin}, {Viallet}, {Campbell},
  {Lattanzio}  \& {Moc{\'a}k}}{{Arnett} et~al.}{2015}]{Arnett_et_al_2015}
{Arnett} W.~D.,  {Meakin} C.,  {Viallet} M.,  {Campbell} S.~W.,  {Lattanzio}
  J.~C.,   {Moc{\'a}k} M.,  2015, \mn@doi [\apj] {10.1088/0004-637X/809/1/30},
  \href {https://ui.adsabs.harvard.edu/abs/2015ApJ...809...30A} {809, 30}

\bibitem[\protect\citeauthoryear{{Aurnou}, {Horn}  \& {Julien}}{{Aurnou}
  et~al.}{2020}]{Aurnou_et_al_2020}
{Aurnou} J.~M.,  {Horn} S.,   {Julien} K.,  2020, \mn@doi [Physical Review
  Research] {10.1103/PhysRevResearch.2.043115}, \href
  {https://ui.adsabs.harvard.edu/abs/2020PhRvR...2d3115A} {2, 043115}

\bibitem[\protect\citeauthoryear{{Backus}}{{Backus}}{1975}]{Backus_1975}
{Backus} G.~E.,  1975, \mn@doi [Proceedings of the National Academy of Science]
  {10.1073/pnas.72.4.1555}, \href
  {https://ui.adsabs.harvard.edu/abs/1975PNAS...72.1555B} {72, 1555}

\bibitem[\protect\citeauthoryear{{Barker}, {Dempsey}  \& {Lithwick}}{{Barker}
  et~al.}{2014}]{Barker_et_al_2014}
{Barker} A.~J.,  {Dempsey} A.~M.,   {Lithwick} Y.,  2014, \mn@doi [\apj]
  {10.1088/0004-637X/791/1/13}, \href
  {https://ui.adsabs.harvard.edu/abs/2014ApJ...791...13B} {791, 13}

\bibitem[\protect\citeauthoryear{{Batygin} \& {Stevenson}}{{Batygin} \&
  {Stevenson}}{2010}]{Batygin_Stevenson_2010}
{Batygin} K.,  {Stevenson} D.~J.,  2010, \mn@doi [\apjl]
  {10.1088/2041-8205/714/2/L238}, \href
  {https://ui.adsabs.harvard.edu/abs/2010ApJ...714L.238B} {714, L238}

\bibitem[\protect\citeauthoryear{{Braginsky} \& {Roberts}}{{Braginsky} \&
  {Roberts}}{1995}]{BraginskyRoberst95}
{Braginsky} S.~I.,  {Roberts} P.~H.,  1995, \mn@doi [Geophysical and
  Astrophysical Fluid Dynamics] {10.1080/03091929508228992}, \href
  {https://ui.adsabs.harvard.edu/abs/1995GApFD..79....1B} {79, 1}

\bibitem[\protect\citeauthoryear{{Browning}, {Brun}  \& {Toomre}}{{Browning}
  et~al.}{2004}]{Browning_etal2004}
{Browning} M.~K.,  {Brun} A.~S.,   {Toomre} J.,  2004, \mn@doi [\apj]
  {10.1086/380198}, \href
  {https://ui.adsabs.harvard.edu/abs/2004ApJ...601..512B} {601, 512}

\bibitem[\protect\citeauthoryear{{Browning}, {Weber}, {Chabrier}  \&
  {Massey}}{{Browning} et~al.}{2016}]{Browning_et_al_2016}
{Browning} M.~K.,  {Weber} M.~A.,  {Chabrier} G.,   {Massey} A.~P.,  2016,
  \mn@doi [\apj] {10.3847/0004-637X/818/2/189}, \href
  {https://ui.adsabs.harvard.edu/abs/2016ApJ...818..189B} {818, 189}

\bibitem[\protect\citeauthoryear{{Brun} \& {Browning}}{{Brun} \&
  {Browning}}{2017}]{Brun_Browning_2017}
{Brun} A.~S.,  {Browning} M.~K.,  2017, \mn@doi [Living Reviews in Solar
  Physics] {10.1007/s41116-017-0007-8}, \href
  {https://ui.adsabs.harvard.edu/abs/2017LRSP...14....4B} {14, 4}

\bibitem[\protect\citeauthoryear{{Burns}, {Vasil}, {Oishi}, {Lecoanet}  \&
  {Brown}}{{Burns} et~al.}{2020}]{Burns_et_al_2020}
{Burns} K.~J.,  {Vasil} G.~M.,  {Oishi} J.~S.,  {Lecoanet} D.,   {Brown} B.~P.,
   2020, \mn@doi [Physical Review Research] {10.1103/PhysRevResearch.2.023068},
  \href {https://ui.adsabs.harvard.edu/abs/2020PhRvR...2b3068B} {2, 023068}

\bibitem[\protect\citeauthoryear{{Canuto}}{{Canuto}}{1997}]{Canuto_1997}
{Canuto} V.~M.,  1997, \mn@doi [\apjl] {10.1086/310955}, \href
  {https://ui.adsabs.harvard.edu/abs/1997ApJ...489L..71C} {489, L71}

\bibitem[\protect\citeauthoryear{{Chandrasekhar}}{{Chandrasekhar}}{1967}]{Chandrasekhar67}
{Chandrasekhar} S.,  1967, {An introduction to the study of stellar structure}

\bibitem[\protect\citeauthoryear{{Chavanne}, {Chill{\`a}}, {Castaing},
  {H{\'e}bral}, {Chabaud}  \& {Chaussy}}{{Chavanne}
  et~al.}{1997}]{chavanne_etal1997}
{Chavanne} X.,  {Chill{\`a}} F.,  {Castaing} B.,  {H{\'e}bral} B.,  {Chabaud}
  B.,   {Chaussy} J.,  1997, \mn@doi [\prl] {10.1103/PhysRevLett.79.3648},
  \href {https://ui.adsabs.harvard.edu/abs/1997PhRvL..79.3648C} {79, 3648}

\bibitem[\protect\citeauthoryear{{Currie} \& {Browning}}{{Currie} \&
  {Browning}}{2017}]{CurrieBrowning17}
{Currie} L.~K.,  {Browning} M.~K.,  2017, \mn@doi [\apj]
  {10.3847/2041-8213/aa8301}, \href
  {https://ui.adsabs.harvard.edu/abs/2017ApJ...845L..17C} {845, L17}

\bibitem[\protect\citeauthoryear{{Currie}, {Barker}, {Lithwick}  \&
  {Browning}}{{Currie} et~al.}{2020}]{Currie_et_al_2020}
{Currie} L.~K.,  {Barker} A.~J.,  {Lithwick} Y.,   {Browning} M.~K.,  2020,
  \mn@doi [\mnras] {10.1093/mnras/staa372}, \href
  {https://ui.adsabs.harvard.edu/abs/2020MNRAS.493.5233C} {493, 5233}

\bibitem[\protect\citeauthoryear{{Featherstone} \& {Hindman}}{{Featherstone} \&
  {Hindman}}{2016}]{Featherstone_Hindman_2016}
{Featherstone} N.~A.,  {Hindman} B.~W.,  2016, \mn@doi [\apj]
  {10.3847/0004-637X/818/1/32}, \href
  {https://ui.adsabs.harvard.edu/abs/2016ApJ...818...32F} {818, 32}

\bibitem[\protect\citeauthoryear{{Gastine}, {Wicht}  \& {Aubert}}{{Gastine}
  et~al.}{2016}]{Gastine_et_al_2016}
{Gastine} T.,  {Wicht} J.,   {Aubert} J.,  2016, \mn@doi [Journal of Fluid
  Mechanics] {10.1017/jfm.2016.659}, \href
  {https://ui.adsabs.harvard.edu/abs/2016JFM...808..690G} {808, 690}

\bibitem[\protect\citeauthoryear{{Gilman}}{{Gilman}}{1978}]{Gilman78}
{Gilman} P.~A.,  1978, \mn@doi [Geophysical and Astrophysical Fluid Dynamics]
  {10.1080/03091927808242661}, \href
  {https://ui.adsabs.harvard.edu/abs/1978GApFD..11..157G} {11, 157}

\bibitem[\protect\citeauthoryear{{Ginet}}{{Ginet}}{1994}]{Ginet_1994}
{Ginet} G.~P.,  1994, \mn@doi [\apj] {10.1086/174374}, \href
  {https://ui.adsabs.harvard.edu/abs/1994ApJ...429..899G} {429, 899}

\bibitem[\protect\citeauthoryear{{Gough} \& {Weiss}}{{Gough} \&
  {Weiss}}{1976}]{Gough_Weiss_1976}
{Gough} D.~O.,  {Weiss} N.~O.,  1976, \mn@doi [\mnras]
  {10.1093/mnras/176.3.589}, \href
  {https://ui.adsabs.harvard.edu/abs/1976MNRAS.176..589G} {176, 589}

\bibitem[\protect\citeauthoryear{{Grossmann} \& {Lohse}}{{Grossmann} \&
  {Lohse}}{2000}]{Grossmann_Lohse_2000}
{Grossmann} S.,  {Lohse} D.,  2000, \mn@doi [Journal of Fluid Mechanics]
  {10.1017/S0022112099007545}, \href
  {https://ui.adsabs.harvard.edu/abs/2000JFM...407...27G} {407, 27}

\bibitem[\protect\citeauthoryear{{Hewitt}, {McKenzie}  \& {Weiss}}{{Hewitt}
  et~al.}{1975}]{Hewitt75}
{Hewitt} J.~M.,  {McKenzie} D.~P.,   {Weiss} N.~O.,  1975, \mn@doi [Journal of
  Fluid Mechanics] {10.1017/S002211207500119X}, \href
  {https://ui.adsabs.harvard.edu/abs/1975JFM....68..721H} {68, 721}

\bibitem[\protect\citeauthoryear{{Hindman}, {Featherstone}  \&
  {Julien}}{{Hindman} et~al.}{2020}]{hindman_etal2020}
{Hindman} B.~W.,  {Featherstone} N.~A.,   {Julien} K.,  2020, \mn@doi [\apj]
  {10.3847/1538-4357/ab9ec2}, \href
  {https://ui.adsabs.harvard.edu/abs/2020ApJ...898..120H} {898, 120}

\bibitem[\protect\citeauthoryear{{Hurlburt}, {Toomre}  \&
  {Massaguer}}{{Hurlburt} et~al.}{1984}]{Hurlburt_et_al_1984}
{Hurlburt} N.~E.,  {Toomre} J.,   {Massaguer} J.~M.,  1984, \mn@doi [\apj]
  {10.1086/162235}, \href
  {https://ui.adsabs.harvard.edu/abs/1984ApJ...282..557H} {282, 557}

\bibitem[\protect\citeauthoryear{{Ireland} \& {Browning}}{{Ireland} \&
  {Browning}}{2018}]{Ireland_Browning_2018}
{Ireland} L.~G.,  {Browning} M.~K.,  2018, \mn@doi [\apj]
  {10.3847/1538-4357/aab3da}, \href
  {https://ui.adsabs.harvard.edu/abs/2018ApJ...856..132I} {856, 132}

\bibitem[\protect\citeauthoryear{{Jarvis} \& {McKenzie}}{{Jarvis} \&
  {McKenzie}}{1980}]{JarvisMcKenzie80}
{Jarvis} G.~T.,  {McKenzie} D.~P.,  1980, \mn@doi [Journal of Fluid Mechanics]
  {10.1017/S002211208000225X}, \href
  {https://ui.adsabs.harvard.edu/abs/1980JFM....96..515J} {96, 515}

\bibitem[\protect\citeauthoryear{{Jermyn}, {Anders}, {Lecoanet}  \&
  {Cantiello}}{{Jermyn} et~al.}{2022}]{Jermyn_et_al_2022}
{Jermyn} A.~S.,  {Anders} E.~H.,  {Lecoanet} D.,   {Cantiello} M.,  2022,
  \mn@doi [\apjs] {10.3847/1538-4365/ac7cee}, \href
  {https://ui.adsabs.harvard.edu/abs/2022ApJS..262...19J} {262, 19}

\bibitem[\protect\citeauthoryear{{Jones}, {Mizerski}  \& {Kessar}}{{Jones}
  et~al.}{2022}]{Jones_et_al_2022}
{Jones} C.~A.,  {Mizerski} K.~A.,   {Kessar} M.,  2022, \mn@doi [Journal of
  Fluid Mechanics] {10.1017/jfm.2021.905}, \href
  {https://ui.adsabs.harvard.edu/abs/2022JFM...930A..13J} {930, A13}

\bibitem[\protect\citeauthoryear{{Julien}, {Knobloch}, {Rubio}  \&
  {Vasil}}{{Julien} et~al.}{2012}]{Julien_et_al_2012}
{Julien} K.,  {Knobloch} E.,  {Rubio} A.~M.,   {Vasil} G.~M.,  2012, \mn@doi
  [\prl] {10.1103/PhysRevLett.109.254503}, \href
  {https://ui.adsabs.harvard.edu/abs/2012PhRvL.109y4503J} {109, 254503}

\bibitem[\protect\citeauthoryear{{Kaspi} et~al.,}{{Kaspi}
  et~al.}{2018}]{Kaspi_et_al_2018}
{Kaspi} Y.,  et~al., 2018, \mn@doi [\nat] {10.1038/nature25793}, \href
  {https://ui.adsabs.harvard.edu/abs/2018Natur.555..223K} {555, 223}

\bibitem[\protect\citeauthoryear{{Kaspi}, {Galanti}, {Showman}, {Stevenson},
  {Guillot}, {Iess}  \& {Bolton}}{{Kaspi} et~al.}{2020}]{Kaspi_et_al_2020}
{Kaspi} Y.,  {Galanti} E.,  {Showman} A.~P.,  {Stevenson} D.~J.,  {Guillot} T.,
   {Iess} L.,   {Bolton} S.~J.,  2020, \mn@doi [\ssr]
  {10.1007/s11214-020-00705-7}, \href
  {https://ui.adsabs.harvard.edu/abs/2020SSRv..216...84K} {216, 84}

\bibitem[\protect\citeauthoryear{{Kazemi}, {Ostilla-M{\'o}nico}  \&
  {Goluskin}}{{Kazemi} et~al.}{2022}]{Kazemi_et_al_2022}
{Kazemi} S.,  {Ostilla-M{\'o}nico} R.,   {Goluskin} D.,  2022, \mn@doi [\prl]
  {10.1103/PhysRevLett.129.024501}, \href
  {https://ui.adsabs.harvard.edu/abs/2022PhRvL.129b4501K} {129, 024501}

\bibitem[\protect\citeauthoryear{{King}, {Stellmach}, {Noir}, {Hansen}  \&
  {Aurnou}}{{King} et~al.}{2009}]{king_etal2009}
{King} E.~M.,  {Stellmach} S.,  {Noir} J.,  {Hansen} U.,   {Aurnou} J.~M.,
  2009, \mn@doi [\nat] {10.1038/nature07647}, \href
  {https://ui.adsabs.harvard.edu/abs/2009Natur.457..301K} {457, 301}

\bibitem[\protect\citeauthoryear{{King}, {Stellmach}  \& {Buffett}}{{King}
  et~al.}{2013}]{King_et_al_2013}
{King} E.~M.,  {Stellmach} S.,   {Buffett} B.,  2013, \mn@doi [Journal of Fluid
  Mechanics] {10.1017/jfm.2012.586}, \href
  {https://ui.adsabs.harvard.edu/abs/2013JFM...717..449K} {717, 449}

\bibitem[\protect\citeauthoryear{{Kulsrud}}{{Kulsrud}}{2005}]{Kulsrud_2005}
{Kulsrud} R.~M.,  2005, {Plasma Physics for Astrophysics}

\bibitem[\protect\citeauthoryear{{Kupka}, {Ahlborn}  \& {Weiss}}{{Kupka}
  et~al.}{2022}]{Kupka_et_al_2022}
{Kupka} F.,  {Ahlborn} F.,   {Weiss} A.,  2022, \mn@doi [\aap]
  {10.1051/0004-6361/202243125}, \href
  {https://ui.adsabs.harvard.edu/abs/2022A&A...667A..96K} {667, A96}

\bibitem[\protect\citeauthoryear{{Lantz}}{{Lantz}}{1992}]{Lantz92}
{Lantz} S.~R.,  1992, PhD thesis, CORNELL UNIVERSITY.

\bibitem[\protect\citeauthoryear{{Lecoanet}, {Brown}, {Zweibel}, {Burns},
  {Oishi}  \& {Vasil}}{{Lecoanet} et~al.}{2014}]{Lecoanet14}
{Lecoanet} D.,  {Brown} B.~P.,  {Zweibel} E.~G.,  {Burns} K.~J.,  {Oishi}
  J.~S.,   {Vasil} G.~M.,  2014, \mn@doi [\apj] {10.1088/0004-637X/797/2/94},
  \href {https://ui.adsabs.harvard.edu/abs/2014ApJ...797...94L} {797, 94}

\bibitem[\protect\citeauthoryear{{Liu}, {Goldreich}  \& {Stevenson}}{{Liu}
  et~al.}{2008}]{Liu_et_al_2008}
{Liu} J.,  {Goldreich} P.~M.,   {Stevenson} D.~J.,  2008, \mn@doi [\icarus]
  {10.1016/j.icarus.2007.11.036}, \href
  {https://ui.adsabs.harvard.edu/abs/2008Icar..196..653L} {196, 653}

\bibitem[\protect\citeauthoryear{Long, Mound, Davies  \& Tobias}{Long
  et~al.}{2020a}]{Long_et_al_2020}
Long R.~S.,  Mound J.~E.,  Davies C.~J.,   Tobias S.~M.,  2020a, \mn@doi [Phys.
  Rev. Fluids] {10.1103/PhysRevFluids.5.113502}, 5, 113502

\bibitem[\protect\citeauthoryear{{Long}, {Mound}, {Davies}  \& {Tobias}}{{Long}
  et~al.}{2020b}]{long_etal2020_scaling}
{Long} R.~S.,  {Mound} J.~E.,  {Davies} C.~J.,   {Tobias} S.~M.,  2020b,
  \mn@doi [Journal of Fluid Mechanics] {10.1017/jfm.2020.67}, \href
  {https://ui.adsabs.harvard.edu/abs/2020JFM...889A...7L} {889, A7}

\bibitem[\protect\citeauthoryear{{Malkus}}{{Malkus}}{1954}]{Malkus_1954}
{Malkus} W.~V.~R.,  1954, \mn@doi [Proceedings of the Royal Society of London
  Series A] {10.1098/rspa.1954.0197}, \href
  {https://ui.adsabs.harvard.edu/abs/1954RSPSA.225..196M} {225, 196}

\bibitem[\protect\citeauthoryear{{Meakin} \& {Arnett}}{{Meakin} \&
  {Arnett}}{2010}]{Meakin_Arnett_2010}
{Meakin} C.~A.,  {Arnett} W.~D.,  2010, \mn@doi [\apss]
  {10.1007/s10509-010-0301-6}, \href
  {https://ui.adsabs.harvard.edu/abs/2010Ap&SS.328..221M} {328, 221}

\bibitem[\protect\citeauthoryear{{Mizerski} \& {Tobias}}{{Mizerski} \&
  {Tobias}}{2011}]{Miserski_Tobias_2011}
{Mizerski} K.~A.,  {Tobias} S.~M.,  2011, \mn@doi [Geophysical and
  Astrophysical Fluid Dynamics] {10.1080/03091929.2010.521748}, \href
  {https://ui.adsabs.harvard.edu/abs/2011GApFD.105..566M} {105, 566}

\bibitem[\protect\citeauthoryear{{Nicoski}, {O'Connor}  \& {Calkins}}{{Nicoski}
  et~al.}{2023}]{nicoski_connor_calkins2023}
{Nicoski} J.~A.,  {O'Connor} A.~R.,   {Calkins} M.~A.,  2023, \mn@doi [arXiv
  e-prints] {10.48550/arXiv.2308.05174}, \href
  {https://ui.adsabs.harvard.edu/abs/2023arXiv230805174N} {p. arXiv:2308.05174}

\bibitem[\protect\citeauthoryear{{Oishi}, {Burns}, {Clark}, {Anders}, {Brown},
  {Vasil}  \& {Lecoanet}}{{Oishi} et~al.}{2021}]{oishi_etal2021_eigentools}
{Oishi} J.,  {Burns} K.,  {Clark} S.,  {Anders} E.,  {Brown} B.,  {Vasil} G.,
  {Lecoanet} D.,  2021, \mn@doi [The Journal of Open Source Software]
  {10.21105/joss.03079}, \href
  {https://ui.adsabs.harvard.edu/abs/2021JOSS....6.3079O} {6, 3079}

\bibitem[\protect\citeauthoryear{{Priest}}{{Priest}}{2014}]{priest2014_book}
{Priest} E.,  2014, {Magnetohydrodynamics of the Sun},
  \mn@doi{10.1017/CBO9781139020732.
}

\bibitem[\protect\citeauthoryear{Rogers, Glatzmaier  \& Woosley}{Rogers
  et~al.}{2003}]{Rogers_et_al_2003}
Rogers T.~M.,  Glatzmaier G.~A.,   Woosley S.~E.,  2003, \mn@doi [Phys. Rev. E]
  {10.1103/PhysRevE.67.026315}, 67, 026315

\bibitem[\protect\citeauthoryear{{Shraiman} \& {Siggia}}{{Shraiman} \&
  {Siggia}}{2000}]{Shraiman_Siggia_2000}
{Shraiman} B.~I.,  {Siggia} E.~D.,  2000, \mn@doi [\nat] {10.1038/35015000},
  \href {https://ui.adsabs.harvard.edu/abs/2000Natur.405..639S} {405, 639}

\bibitem[\protect\citeauthoryear{{Siggia}}{{Siggia}}{1994}]{Siggia_1994}
{Siggia} E.~D.,  1994, \mn@doi [Annual Review of Fluid Mechanics]
  {10.1146/annurev.fl.26.010194.001033}, \href
  {https://ui.adsabs.harvard.edu/abs/1994AnRFM..26..137S} {26, 137}

\bibitem[\protect\citeauthoryear{{Stein} \& {Nordlund}}{{Stein} \&
  {Nordlund}}{1989}]{Stein_Nordlund_1989}
{Stein} R.~F.,  {Nordlund} A.,  1989, \mn@doi [\apjl] {10.1086/185493}, \href
  {https://ui.adsabs.harvard.edu/abs/1989ApJ...342L..95S} {342, L95}

\bibitem[\protect\citeauthoryear{{Stevenson}}{{Stevenson}}{1979}]{Stevenson_1979}
{Stevenson} D.~J.,  1979, \mn@doi [Geophysical and Astrophysical Fluid
  Dynamics] {10.1080/03091927908242681}, \href
  {https://ui.adsabs.harvard.edu/abs/1979GApFD..12..139S} {12, 139}

\bibitem[\protect\citeauthoryear{{Vasil}, {Julien}  \& {Featherstone}}{{Vasil}
  et~al.}{2021}]{Vasil_et_al_2021}
{Vasil} G.~M.,  {Julien} K.,   {Featherstone} N.~A.,  2021, \mn@doi
  [Proceedings of the National Academy of Science] {10.1073/pnas.2022518118},
  \href {https://ui.adsabs.harvard.edu/abs/2021PNAS..11822518V} {118,
  e2022518118}

\bibitem[\protect\citeauthoryear{{Viallet}, {Meakin}, {Arnett}  \&
  {Moc{\'a}k}}{{Viallet} et~al.}{2013}]{Viallet13}
{Viallet} M.,  {Meakin} C.,  {Arnett} D.,   {Moc{\'a}k} M.,  2013, \mn@doi
  [\apj] {10.1088/0004-637X/769/1/1}, \href
  {https://ui.adsabs.harvard.edu/abs/2013ApJ...769....1V} {769, 1}

\bibitem[\protect\citeauthoryear{{Wang}, {Chong}, {Stevens}, {Verzicco}  \&
  {Lohse}}{{Wang} et~al.}{2020}]{wang_etal2020}
{Wang} Q.,  {Chong} K.~L.,  {Stevens} R. J.~A.~M.,  {Verzicco} R.,   {Lohse}
  D.,  2020, \mn@doi [Journal of Fluid Mechanics] {10.1017/jfm.2020.793}, \href
  {https://ui.adsabs.harvard.edu/abs/2020JFM...905A..21W} {905, A21}

\makeatother
\end{thebibliography}








\bsp	
\label{lastpage}
\end{document}